 \documentclass[prl,amsmath,amssymb,twocolumn, showpacs, superscriptaddress,10pt]{revtex4-1} 

\usepackage{amsmath}
\usepackage{hyperref}
\usepackage{graphicx}
\usepackage{amsfonts}
\usepackage{amsthm}
\usepackage{cases}
\usepackage{bm}
\usepackage[utf8]{inputenc}
\usepackage{enumitem}
\usepackage[caption=false]{subfig}

\usepackage[normalem]{ulem}

\usepackage{color}
\definecolor{Blue}{rgb}{0.00, 0.00, 1.00}
\definecolor{Red}{rgb}{1.00, 0.00, 0.00}
\definecolor{Green}{rgb}{0.00, 0.70, 0.00}

\def\XXint#1#2#3{{\setbox0=\hbox{$#1{#2#3}{\int}$}
     \vcenter{\hbox{$#2#3$}}\kern-.5\wd0}}

\usepackage{ifxetex}
\ifxetex
\usepackage{fontspec}
    \defaultfontfeatures{Ligatures={TeX}}
\usepackage[math-style=ISO,bold-style=ISO]{unicode-math}
\else
    \usepackage{esint}
\fi

\hypersetup{
    colorlinks=true,       
    linkcolor=red,          
    citecolor=blue,        
    filecolor=magenta,      
    urlcolor=cyan           
}

\newcommand{\nn}{\nonumber}
\newcommand{\be}{\begin{equation}}
\newcommand{\ee}{\end{equation}}
\newcommand{\bea}{\begin{eqnarray}}
\newcommand{\eea}{\end{eqnarray}}

\newcommand{\beq}{\begin{equation}}
\newcommand{\eeq}{\end{equation}}
\newcommand{\beqn}{\begin{eqnarray}}
\newcommand{\eeqn}{\end{eqnarray}}

\DeclareMathOperator{\sgn}{sgn}

\begin{document}

\title{Non-equilibrium phase transitions in active rank diffusions}

\author{Léo Touzo}
\affiliation{Laboratoire de Physique de l'\'Ecole Normale Sup\'erieure, CNRS, ENS $\&$ PSL University, Sorbonne Universit\'e, Universit\'e Paris Cité, 75005 Paris, France}
\author{Pierre Le Doussal}
\affiliation{Laboratoire de Physique de l'\'Ecole Normale Sup\'erieure, CNRS, ENS $\&$ PSL University, Sorbonne Universit\'e, Universit\'e Paris Cité, 75005 Paris, France}

\date{\today}

\begin{abstract}
We consider $N$ run and tumble particles in one dimension interacting via 
a linear 1D Coulomb potential,
an active version of the rank diffusion problem. It was solved previously for $N=2$ 
leading to a stationary bound state in the attractive case. Here the evolution of the density fields is obtained in the large $N$ limit 
in terms of two coupled Burger's type equations. In the attractive case the exact stationary solution
describes a non-trivial $N$-particle bound state, which exhibits transitions between a phase 
where the density is smooth with infinite support, a phase where the density has finite support and exhibits "shocks", i.e. clusters of particles, 
at the edges, and a fully clustered phase.
In presence of an additional linear potential, the phase diagram, obtained for either sign of the interaction,
is even richer, with additional partially expanding phases, with or without shocks.
Finally, a general self-consistent method is introduced to treat 
more general interactions.
The predictions are tested 
through extensive numerical simulations. 
\end{abstract}


\maketitle


The run-and-tumble particle (RTP),
driven by telegraphic noise \cite{ML17,kac74},
is the simplest model of an active
particle which mimics e.g. the motion of E. Coli bacteria \cite{Berg2004,TailleurCates}.
Interacting RTP's, which exhibit remarkable collective effects,
is a thriving area of research 
\cite{TailleurCates,FM2012,soft,CT2015,slowman,slowman2,BG2021,SG2014}. One outstanding question is to characterize the 
non Gibbsian steady state reached at large time.
Even in one dimension (1D), there are very few exact results, beyond two interacting RTP's \cite{slowman,slowman2,us_bound_state,Maes_bound_state,nonexistence,MBE2019,KunduGap2020,LMS2019}, or for 
harmonic chains \cite{SinghChain2020,PutBerxVanderzande2019}. Recently, analytical results were obtained for some many-particle models on a 1D lattice with contact interactions~\cite{Metson2022,MetsonLong,Dandekar2020,Thom2011}, and for an active version 
of the Dyson Brownian motion in the continuum with logarithmic interactions \cite{TouzoDBM2023}. 

An analytically tractable {\it passive} example 
is the Langevin dynamics of Brownian particles interacting 
via the linear Coulomb 1D potential. It is known as {\it ranked diffusion}
since the force acting on each particle is proportional to its rank, i.e., the number of particles in front of it, 
minus the number 
behind. Introduced in finance and mathematics 
\cite{Banner,Pitman}, it is related to the statistical mechanics of the self-gravitating 1D gas \cite{Rybicki,Sire}
(attractive case) and to the Jellium model (repulsive case, in presence of a confining potential) much studied
at equilibrium 
\cite{Lenard,Baxter,Tellez,Lewin,SatyaJellium1,Chafai_edge,Flack22}. 
Recently, a detailed description of the non-equilibrium dynamics was obtained \cite{PLDRankedDiffusion}, using both
a mapping to the delta Bose gas, solvable by the Bethe ansatz,
as well as the Dean-Kawasaki approach \cite{Dean,Kawa}, which leads to a description of the density
in terms of a noisy Burgers equation \cite{PLDRankedDiffusion}. In the attractive case, the
shock solutions of the Burgers equation correspond to macroscopic bound states,
while in the repulsive case the system forms an expanding crystal at large time
\cite{FlackRD}.

In view of these results, it is natural to extend the model to the active case. 
In this Letter
we study $N$ run-and-tumble particles (RTP) at positions
$x_i(t)$ in $d=1$, interacting
via the 1D Coulomb potential and evolving via the equation of motion
\be \label{langevin1}
\frac{dx_i}{dt} = \frac{\bar \kappa}{N} \sum_{j=1}^N {\rm sgn}(x_j-x_i) - V'(x_i) + v_0 \sigma_i(t) + \sqrt{2 T} \xi_i(t) 
\ee 
Here the $\xi_i(t)$ are unit independent white noises and the $\sigma_i(t)=\pm 1$
are independent telegraphic noises with rate $\gamma$.
The particles can cross freely, i.e.
there is no hard core repulsion, and by convention ${\rm sgn}(0)=0$.
The case $\bar \kappa>0$ corresponds to attractive interactions 
and $\kappa=-\bar \kappa>0$ to repulsive ones. 
The RTP's evolve in a common external potential $V(x)$.
When $V(x)=0$ the center of mass $\bar x=\frac{1}{N} \sum_i x_i$
evolves as a sum of $N$ independent RTP's with velocities $v_0/N$,
hence it diffuses at large time as $\bar x \sim \sqrt{2 D_N t}$ 
with $D_N=\frac{1}{N}(T + \frac{v_0^2}{2 \gamma})$.

Until now this model was studied only for $N=2$, in the attractive case $\bar \kappa>0$, and for $V(x)=0$.
It was shown \cite{us_bound_state} that the PDF $P(z,t)$ of the relative coordinate
$z=x_1-\bar x=\frac{x_1-x_2}{2}$ reaches a stationary limit $P(z)$
which describes a bound state. For $T>0$, $P(z)$ is the sum of three decaying exponentials. 
In the purely active limit $T=0$, some of these exponentials become delta functions, and it was found \cite{us_bound_state} that
for $\frac{v_0}{\bar \kappa}> \frac{1}{2}$ 
\bea  \label{tot_prob1}
&& P(z) =  \frac{1-c_2}{2 \xi_2} e^{- \frac{|z|}{\xi_2}} + c_2 \delta(z) 
\\
&& \xi_2= \frac{4 v_0^2 - \bar \kappa^2}{8 \gamma \bar \kappa} 
~ , ~ c_2=
\frac{\bar \kappa^2}{4 v_0^2 + \bar \kappa^2}
 \;,
\eea
while for $\frac{v_0}{\bar \kappa} < \frac{1}{2}$ one has $P(z)=\delta(z)$,
i.e all particles form a single cluster ($c_2=1$). Indeed since particles
at the same position do not interact (${\rm sgn}(0)=0$), 
clusters tend to form when the interaction is strong enough.
Obtaining the full stationary distribution for any number $N>2$ of particles would be
very interesting, but is quite challenging \cite{PLDGS}. Here we
study the time dependent density fields $\rho_\sigma(x,t)$ with $\sigma= \pm 1$
and their
even and odd components $\rho_s$ and $\rho_d$, defined as
\bea
&& \rho_\sigma(x,t) = \frac{1}{N} \sum_i \delta(x_i(t)-x) \delta_{\sigma_i(t),\sigma} \\
&& \rho_{s/d}(x,t)=\rho_+(x,t) \pm \rho_-(x,t) \,. 
\eea 
An outstanding question is whether the delta peak in the density survives when
$N$ increases. We first investigate this question numerically for finite $N$. 
Next, using the Dean-Kawasaki approach \cite{Dean,Kawa}, as in 
\cite{PLDRankedDiffusion} for the passive case, and in \cite{TouzoDBM2023} 
for the active Dyson Brownian motion, we show that the density fields evolve according to 
two coupled stochastic non-linear equations. In the large $N$
limit these equations become deterministic and of the Burgers type. From them
we obtain the exact solution for the $N$ particle stationary bound 
state. We also extend our results in the presence of an external
linear potential both in the attractive and repulsive case.
This leads to a rich phase diagram at $T=0$, with phases where the 
density is either smooth or contains shocks (i.e clusters)
at various positions, as well as phases describing
an active expanding crystal.

\begin{figure}
    \centering
    \includegraphics[width=0.49\linewidth]{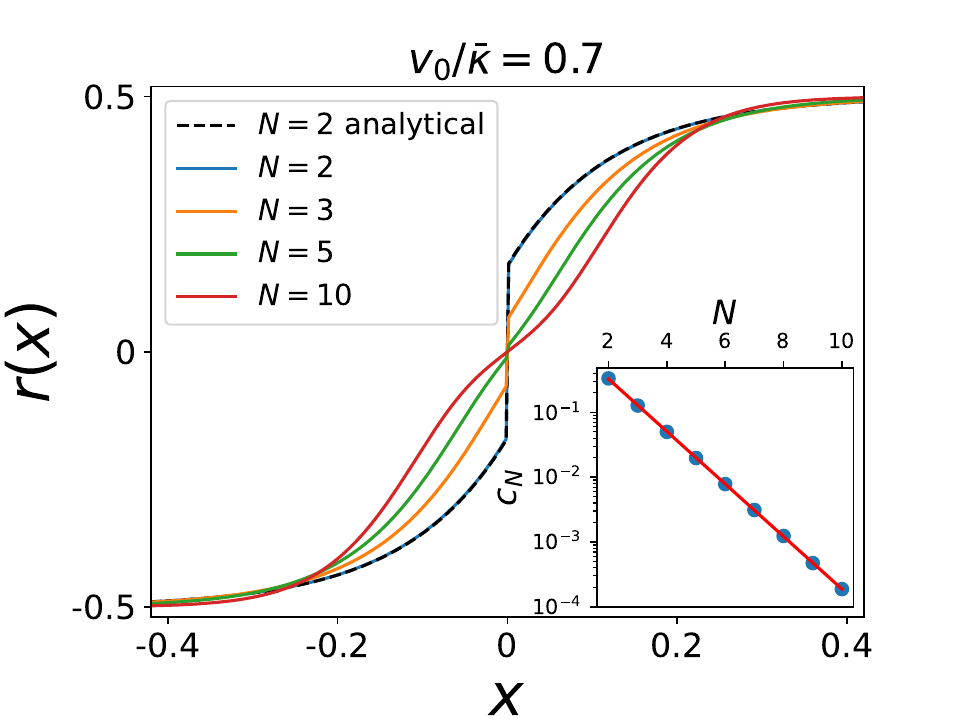}
    \includegraphics[width=0.49\linewidth]{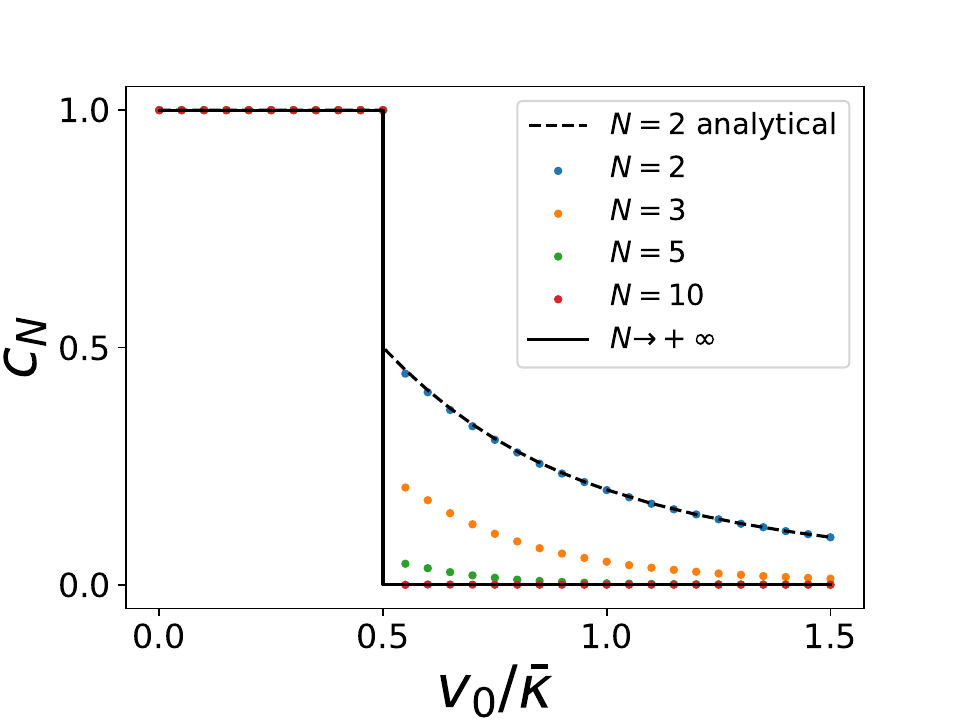}
    \caption{{\bf Left:} Rank field $r(x)=\int_{-\infty}^x dy \, \rho_s(y) -\frac{1}{2}$ in the stationary state, obtained through simulations, as a function of $x$, in the attractive case with $v_0=0.7$, $\bar \kappa=1$ and $\gamma=1$, for small values of $N$. The analytical prediction 
    from \eqref{tot_prob1} for $N=2$ is also shown. The density has a delta peak at $x=0$, i.e. a jump in $r(x)$,
    with a weight $c_N$, which decreases exponentially with $N$ (see inset). {\bf Right:} Weight $c_N$ as a function of $v_0/\bar \kappa$ for different values of $N$. The black line corresponds to the limit $N \to+\infty$ (see below).}
    \label{fig_smallN}
\end{figure}

We have performed numerical simulations of Eq. \eqref{langevin1} at $T=0$
for various numbers of particles $N$. In the attractive case with $V(x)=0$ we have measured the
density of particles in the stationary state by performing time averages over a long time window.
Here the density is measured in the reference frame of the instantaneous center of mass
(this distinction becomes irrelevant
at large $N$). The results are
shown in Fig. \ref{fig_smallN}. 
One finds that the delta function at $x=0$ in $\rho_s(x)$ 
persists for any finite $N$, with an amplitude $c_N$ which depends
on $\frac{v_0}{\bar \kappa}$ and on $N$ (which is the only dimensionless parameter
in that case). The first result is that for $\frac{v_0}{\bar \kappa}<1/2$
all particles are in a single cluster at $x=0$, i.e. $c_N=1$ for any $N$.
Indeed, one can show that this configuration is stable.
Denote $x_+$ (resp. $x_-$) the position of one of the rightmost (resp. leftmost) particles
and $n_+$ (resp. $n_-$) the number of particles at the same location.
One has 
\be 
\frac{d(x_+-x_-)}{dt} \leq 2 v_0 - \bar \kappa \left( 2-\frac{n_++n_-}{N} \right)
\leq 2v_0 - \bar \kappa
\ee 
since $n_++n_- \leq N$. Hence for $\frac{v_0}{\bar \kappa}<1/2$ the
width of the support always decreases to $0$. 
For $\frac{v_0}{\bar \kappa}>1/2$ we find that the amplitude $c_N$ 
decreases with $\frac{v_0}{\bar \kappa}$ and 
decays exponentially in $N$. 

\begin{figure}[t]
    \centering
    \includegraphics[width=0.95\linewidth, trim={0.1cm 3.8cm 0.1cm 10cm}, clip]{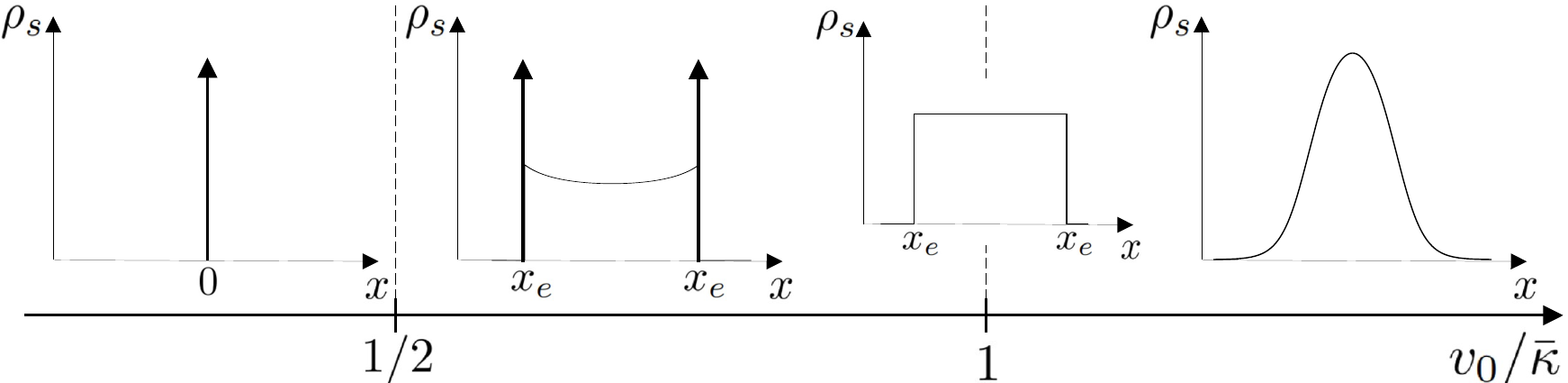}
    \caption{Phase diagram of active ranked diffusion in the attractive case (in the absence of external potential). For each phase (as well as in the marginal case $v_0=\bar \kappa$), the density $\rho_s$ in the $N \to + \infty$ limit is represented, for some values of the parameters (the arrows represent delta functions).}
    \label{phase_diagram}
\end{figure}

A remarkable feature, as we will see
below, is that in the limit $N \to +\infty$ the delta peak
at $x=0$ is absent for $\frac{v_0}{\bar \kappa}>1/2$, but 
for $v_0<\bar \kappa$ it is replaced by a pair of peaks at
positions $\pm x_e$, which mark the edge of the (finite)
support. A genuine phase transition occurs at $v_0=\bar\kappa$, as indicated
in Fig.~\ref{phase_diagram}, beyond which the stationary density is smooth
and the support extends to infinity. 
We now explain how these results are obtained.

Using the Dean-Kawasaki approach  \cite{Dean,Kawa}, one can establish starting from \eqref{langevin1} 
, as in \cite{PLDRankedDiffusion} and \cite{TouzoDBM2023}, an exact stochastic evolution equation
for the density fields, which takes the following form 
\bea \label{eqrho1}
&& \partial_t \rho_\sigma  =  T \partial_x^2 \rho_\sigma  +  \partial_x \rho_\sigma  \big( - v_0 \sigma +  V'(x) \\ 
&& + \bar \kappa \int dy \, \rho_s(y,t) {\rm sgn}(x-y) \big) + \gamma \rho_{- \sigma} - \gamma \rho_{\sigma}
+ O(\frac{1}{\sqrt{N}}) \nonumber
\eea
where the $O(\frac{1}{\sqrt{N}})$ term represents the passive and active noises,
see \cite{SM}, and Sec. III A of \cite{TouzoDBM2023}.
It is convenient to define, as in \cite{PLDRankedDiffusion},
the rank fields
\be \label{rankdef1}
r(x,t) = \int^x_{-\infty} dy \, \rho_s(y,t)\, -\, \frac{1}{2} \; , \; s(x,t) = \int_{-\infty}^x dy \, \rho_d(y,t) \; .
\ee
Focusing from now on on the large $N$ limit, and henceforth neglecting the noise terms in \eqref{eqrho1},
one obtains from \eqref{eqrho1} and \eqref{rankdef1} two coupled deterministic differential equations (see \cite{SM})
\bea \label{system1}
\!\!\!\!\!\!  \partial_t r &=& T \partial_x^2 r - v_0 \partial_x s
+ 2 \bar \kappa r \partial_x r + V'(x) \partial_x r \; , \\
\!\!\!\!\!\!  \partial_t s &=& T  \partial_x^2 s  - v_0 \partial_x r  + 2  \bar \kappa  r \partial_x s + V'(x) \partial_x s - 2 \gamma s \label{system2} \;.
\eea 
These equations are valid both for the attractive and the repulsive case. Since $\rho_s(x,t)$ is positive and normalized to 1, $r(x,t)$ must be an increasing function with $r(\pm \infty,t)=\pm \frac{1}{2}$. One sees that $\partial_t s(+\infty,t)=-2 \gamma s(+\infty,t)$, hence at large time
$s(\pm \infty,t)=0$. 
In the passive case $v_0=0$ the first equation recovers Burger's equation which
describes usual rank diffusion \cite{PLDRankedDiffusion}. 
One generally expects that in the diffusive limit $v_0,\gamma \to +\infty$ with 
$T_a=\frac{v_0^2}{2 \gamma}$ fixed, RTP's behave as Brownian particles. This also holds here. Indeed, in \eqref{system2} only two terms 
remain relevant, and one obtains $s \simeq - \frac{v_0}{2\gamma} \partial_x r$. Inserting in \eqref{system1} the active term 
$- v_0 \partial_x s(x,t)$ becomes $T_a \partial_x^2 r(x,t)$, i.e. a thermal term. 

We now turn to the analysis of the large time limit of these equations, starting with the case $V(x)=0$. 
\\

{\bf Attractive case}. We set here $V(x)=0$. A stationary density then
exists only in the attractive case $\bar \kappa>0$, which we now consider. 
To find this stationary density, we set the
time derivatives to zero in \eqref{system1}, \eqref{system2}. The resulting
equations are invariant by translation, hence we choose coordinates
such that the center of mass $\bar x=\int dx \, x \, r'(x)=0$. 
The first equation can be integrated leading to
\be \label{integr1} 
T r' (x) - v_0 s(x) + \bar  \kappa \big(r(x)^2 - \frac{1}{4}\big) = 0 \;.
\ee

\begin{figure}
    \centering
    \includegraphics[width=0.49\linewidth]{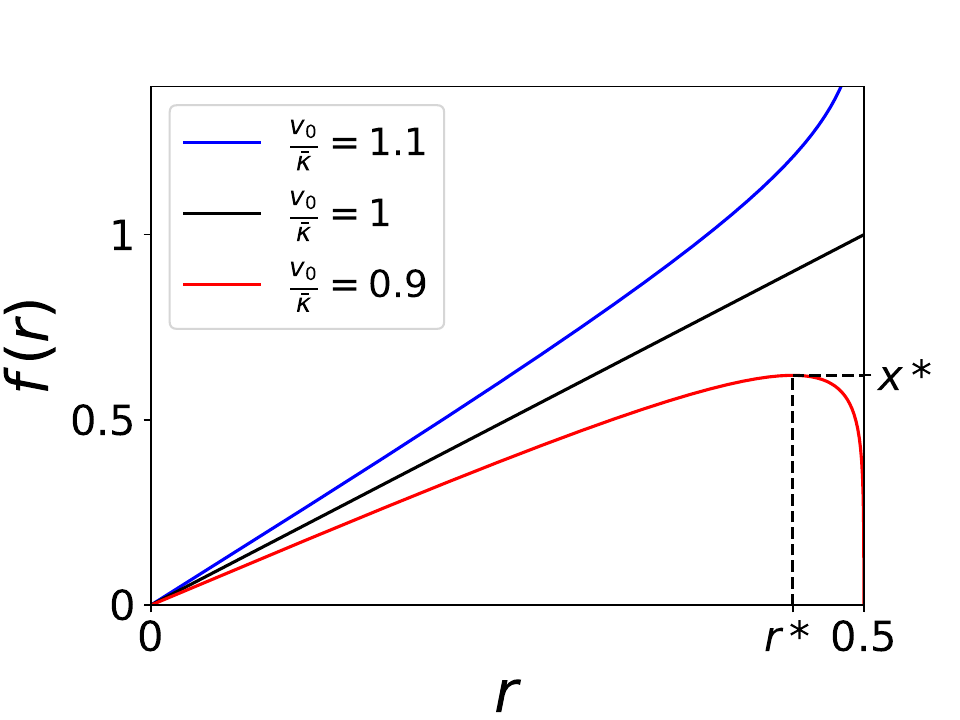}
    \includegraphics[width=0.49\linewidth]{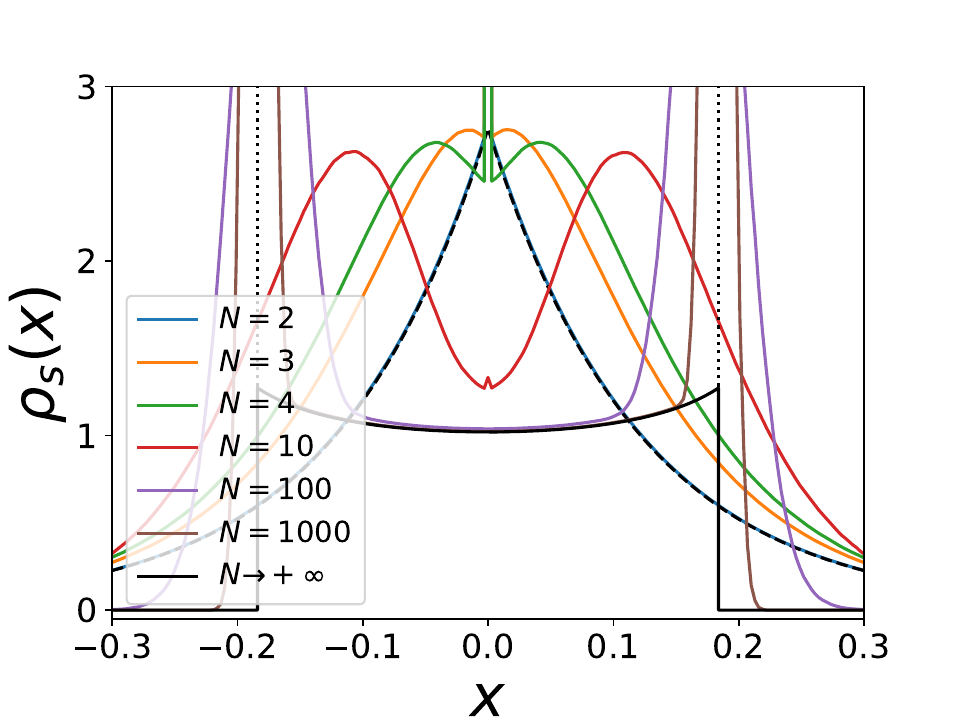}
    \caption{{\bf Left:} The (odd) function $f(r)$ in \eqref{solu0} for $r\geq 0$ and for 3 different values of $\frac{v_0}{\bar \kappa}$. For
    $\frac{v_0}{\bar \kappa}>1$ it diverges at $r=\frac{1}{2}$. For $\frac{v_0}{\bar \kappa}<1$ a maximum appears at $r=r^*<\frac{1}{2}$.
    {\bf Right:} Density $\rho_s(x)$ for $v_0=0.7$, $\bar \kappa=1$ and $\gamma=1$: as $N$ increases it takes
    a bimodal shape, with two smooth symmetric peaks which, for $N=+\infty$, become delta peaks (shocks) at the two edges (shown as dotted vertical lines). The histogram shows schematically the finite $N$ delta peak at $x=0$. The dashed
    black line shows the analytical prediction \eqref{tot_prob1} for $N=2$. The plot of $\rho_s$ for $N\to+\infty$ is obtained using the parametric representation \eqref{rho_parametric} for $a=0$.}
    \label{fig_attractive}
\end{figure}

In the absence of activity, $v_0=0$, the solution is that of standard rank diffusion, 
\cite{PLDRankedDiffusion}
\be \label{soliton} 
r(x)=\frac{1}{2} \tanh\left( \frac{\ \bar \kappa x}{2 T} \right)  \quad , \quad 
\rho_s(x) = \frac{\bar \kappa}{4 T \cosh( \frac{\ \bar \kappa x}{2 T} )^2} \; .
\ee
which for $T \to 0$ becomes a shock solution of Burgers equation,
$r(x)= \frac{1}{2} {\rm sgn}(x)$, i.e 
a single delta peak in the density $\rho_s(x)=r'(x)=\delta(x)$. At $T>0$
this shock solution acquires a finite width $x = O(T)$. One expects that for $v_0>0$
a similar rounding occurs at low $T$, not studied here. Instead, we focus directly on $T=0$ and $v_0>0$, i.e. pure telegraphic noise. 
In that case one must solve 
\bea \label{eqdepart1} 
&& v_0 s'(x)  = 2 \bar \kappa \, r(x) r'(x) \;,  \\
&& v_0 r'(x) = 2 \bar \kappa r(x) s'(x) - 2 \gamma s(x) \;,
\label{eqdepart2} 
\eea 
keeping in mind the possibility
that $r(x)$ may have jumps (see below). The first equation \eqref{eqdepart1} can be integrated, leading to 
\be \label{s_function_r}
s(x) = \frac{\bar \kappa}{v_0} \big(r(x)^2 - \frac{1}{4} \big) \;.
\ee 
Substituting in the second equation one obtains
\be  \label{rp} 
\big(v_0^2 - 4 \bar \kappa^2 r(x)^2\big) r'(x) = 2 \gamma \bar \kappa \big(\frac{1}{4}-r(x)^2 \big)  \;.
\ee 

Since $|r(x)| \leq 1/2$ for all $x$, with $r(\pm \infty)= \pm 1/2$, the r.h.s. is positive and bounded. 
Let us first consider $v_0>\bar \kappa$, in which case $r'(x)$ is bounded from \eqref{rp}. 
Hence there cannot be a shock and the stationary density is
smooth, with an infinite support, and $r(x)$ is obtained by inversion of the equation
\be  \label{solu0} 
\frac{ \gamma x}{\bar \kappa}  =  f(r(x)) \ , \ f(r) = 2 r + \big(\frac{v_0^2}{\bar \kappa^2}- 1 \big) {\rm arctanh}(2 \, r) \;,
\ee 
where we used again the condition $\bar x=0$ to fix the integration constant
\cite{footnoteparity}. One can check that for $v_0>\bar \kappa$, the function $f(r)$ is indeed
invertible, see Fig. \ref{fig_attractive}. The total density $\rho_s(x)=r'(x)$ is even in $x$, 
while $\rho_d(x)=s'(x)$ is odd in $x$, and both decay
for $|x| \to +\infty$ as
\bea \label{asymptotics} 
&& \rho_s(x) \simeq A_s \, e^{- \frac{|x|-x_e^*}{\xi_\infty}}  ~,~ \rho_d(x) \simeq A_d \, {\rm sgn}(x) e^{- \frac{|x|-x_e^*}{\xi_\infty}} \\
&& \xi_\infty =  \frac{v_0^2 - \bar \kappa^2}{ 2 \gamma \bar \kappa  } ~,~ x_e^*=\frac{\bar \kappa}{\gamma}  ~,~  A_s = \frac{1}{\xi_{\infty}}
~ , ~ A_d=  \frac{\bar \kappa}{v_0} A_s \nonumber \;.
\eea 

Upon decreasing $\frac{v_0}{\bar \kappa}$ one sees from \eqref{asymptotics} that
a transition must occur at $\frac{v_0}{\bar \kappa}=1$. Indeed at 
this point $f(r)=2 r$, hence for $\frac{v_0}{\bar \kappa}=1$ the above solution \eqref{solu0} becomes
\be \label{v0=kappa_solr}
r(x) = \frac{1}{2} \frac{x}{x_e^*} ~,~ s(x) = - \frac{1}{4} \big(1 - (\frac{x}{x^*_e})^2 \big) 
~,~ |x| \leq x_e^*
\ee 
and $r(x)=\frac{1}{2} {\rm sgn}(x)$ and $s(x)=0$ for $|x| \geq x_e$. Hence the densities
are non-zero in a finite interval $[-x_e,x_e]$, with 
\be \label{v0=kappa_solrho}
\rho_s(x) = \frac{\gamma}{2 \bar \kappa} ~,~ \rho_d(x) = \frac{ \gamma^2 x}{2 \bar \kappa^2}
~,~ |x| \leq x_e^*= \frac{\bar \kappa}{\gamma} 
\ee 
and vanish for $|x|>x_e^*$, with step discontinuities at the two edges. 

\begin{figure}
    \centering
    \includegraphics[width=0.49\linewidth]{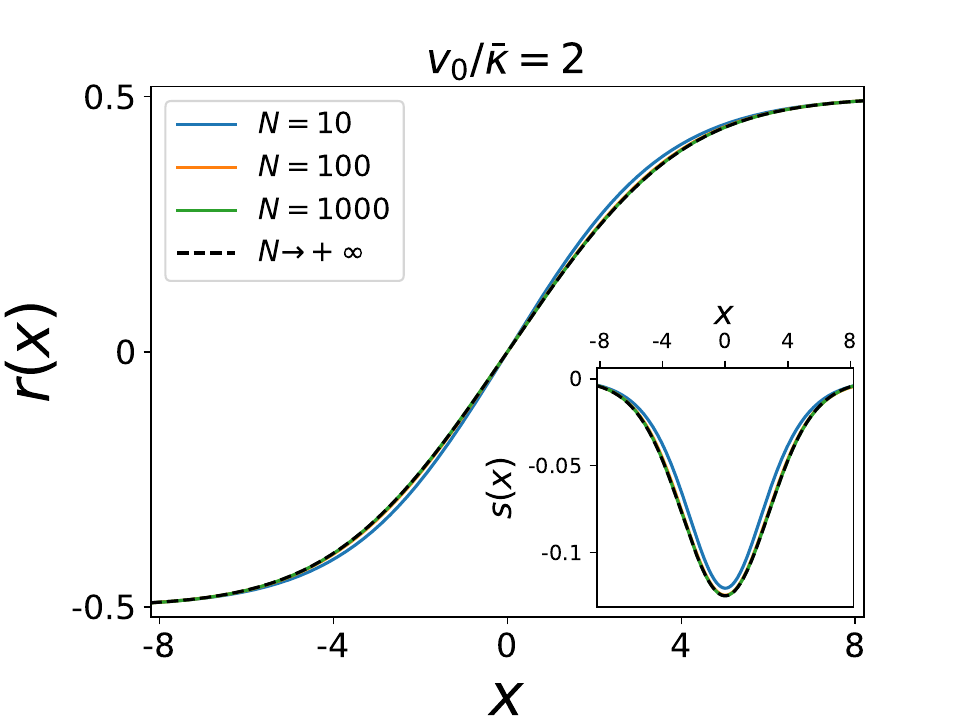}
    \includegraphics[width=0.49\linewidth]{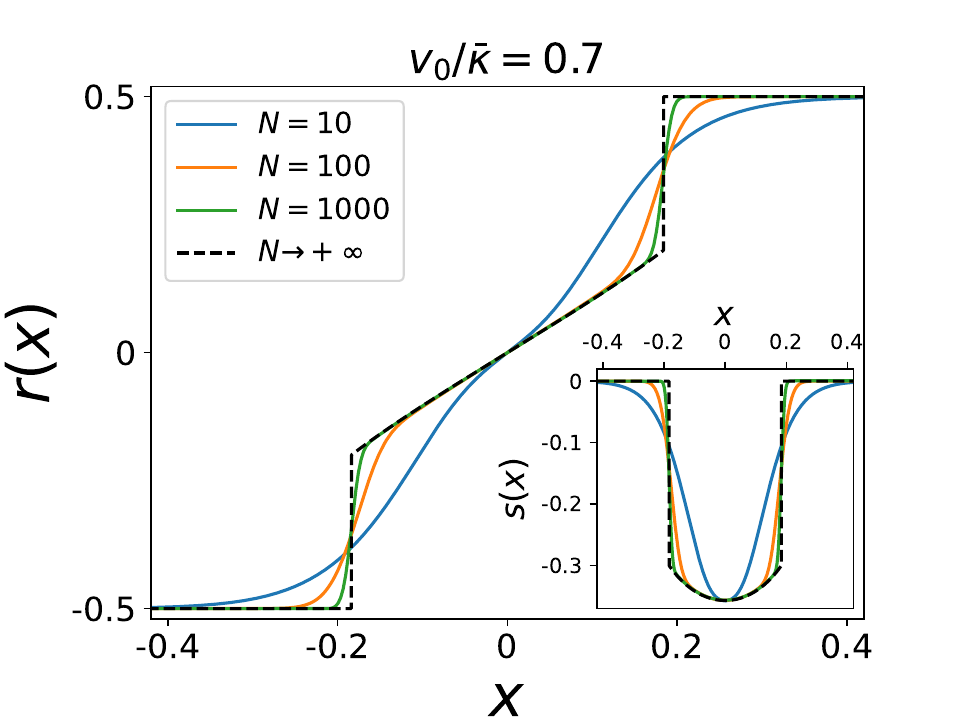}
    \caption{Rank field $r(x)$ in the stationary state, obtained through simulations, as a function of $x$, in the attractive case with $\bar \kappa=1$ and $\gamma=1$, for increasing values of $N$. The dashed black curves show the analytical prediction for $N \to +\infty$. The insets show $s(x)$ for the same set of parameters. {\bf Left:} $v_0=2$. The density is smooth and has infinite support. {\bf Right:} $v_0=0.7$. The density has finite support and exhibits $\delta$ peaks at the edges (i.e jumps in $r(x)$ and $s(x)$). }
    \label{fig_attractive_r}
\end{figure}

For $v_0<\bar \kappa$, $f(r)$ becomes non-invertible, as can be seen from Fig. \ref{fig_attractive}. 
In this case the density has a finite support with edges at $\pm x_e$. More precisely, $f(r)$ is increasing only on an interval $[-r^*,r^*]$ with $r^*=\frac{v_0}{2\bar \kappa}<\frac{1}{2}$. Since $r(x)$ should be increasing, one necessarily has $x_e \leq x^*  = \frac{\bar \kappa}{\gamma} f(r^*)$ and $r(x)$ must have shocks at $\pm x_e$. Since the r.h.s. in \eqref{rp} is bounded, $r'(x)$ can only diverge at a point where the prefactor in the l.h.s of \eqref{rp} vanishes. Naively, this would lead to $r(x_e) = r^*$, i.e. $x_e=x^*$. However, at the position of a shock, the interaction term should be interpreted with care,
as for the standard Burgers equation. Due to the convention ${\rm sgn}(0)=0$, the force exerted on a cluster of particles at position $x_e$ is $\bar \kappa(\frac{1}{2} - r(x_e^+)) - \bar \kappa(r(x_e^-) + \frac{1}{2}) =
- \bar \kappa (r(x_e^+)+r(x_e^-))$. 
Integrating equations (\ref{eqdepart1}-\ref{eqdepart2}) from $x_e^-$ to $x_e^+$ then gives \cite{SM}
\be 
\Delta r = \frac{\bar \kappa}{v_0} (r(x_e^+)+r(x_e^-)) \Delta s ~,~ 
\Delta s = \frac{\bar \kappa}{v_0} (r(x_e^+)+r(x_e^-)) \Delta r \nonumber
\ee 
with $\Delta r=r(x_e^+)-r(x_e^-)$ and similarly for $\Delta s$. A non-zero $\Delta r$ then implies $r(x_e^+)+r(x_e^-)=v_0/\bar \kappa$ leading to
\be \label{eq_edge}
r(x_e^-)=\frac{v_0}{\bar \kappa} -\frac{1}{2} \;.
\ee
since $r(x_e^+)=1/2$. 
For $|x|<x_e$, the prefactor in the l.h.s of \eqref{rp} is strictly positive, and thus $r(x)$ is still given by \eqref{solu0}, while for $|x|>x_e$ one has $r(x)=\frac{1}{2}$. This means that $\rho_s(x)$ has delta peaks at $\pm x_e$ each containing a fraction $1-\frac{v_0}{\bar\kappa}$ of particles. $s(x)$ is still given by \eqref{s_function_r} and thus it also has jumps of amplitude $1-\frac{v_0}{\bar\kappa}$ at $\pm x_e$. This implies that all the particles inside the cluster at $\pm x_e$ have $\sigma=\pm 1$ respectively. 
Note that as $\frac{v_0}{\kappa} \to \frac{1}{2}$, $x_e \to 0$ leading to
a unique delta peak in $\rho_s(x)$ at $x=0$, consistent with the stability argument given above.




The existence of these edge clusters can be understood as follows. If there were no clusters, the rightmost particle would be subject to a force $\bar \kappa (1-\frac{1}{N})$ towards the left. Thus when $\bar \kappa > v_0$, for large enough $N$ it will always move towards the left even if it has an intrinsic velocity $+ v_0$. It will then aggregate with other $+$ particles, until the resulting cluster reaches a fraction $n_c/N:=\frac{1}{2}-r(x_e^-)$ of the total number of particles large enough to be at dynamical equilibrium, i.e. $v_0=\bar \kappa (1-n_c/N)$, from which we recover \eqref{eq_edge}. For finite $N$ however, the size of this cluster will fluctuate, hence its position as well, and therefore we will observe a finite peak in the density instead of a delta function.

These predictions for $r(x)$ and $s(x)$ are compared with numerical simulations in Fig.~\ref{fig_attractive_r}. For $v_0>\bar \kappa$, the agreement is good even at very small values of $N$ ($N\sim 10$). For $\frac{\bar \kappa}{2} < v_0<\bar \kappa$, a distinct
phase is predicted for $N=+\infty$. The numerics for finite $N$ clearly shows precursor
signatures of this phase: as one can see on Fig. \ref{fig_attractive} a bimodal form of the density is already visible for small values of $N>2$. 
Furthermore the agreement on Fig.~\ref{fig_attractive_r} at large $N$ for $r(x)$ and $s(x)$ is also very good.
The behavior near the edges and analytical expression for the moments of the 
particle positions are obtained in \cite{SM}.

\begin{figure}
    \centering
    \includegraphics[width=0.49\linewidth]{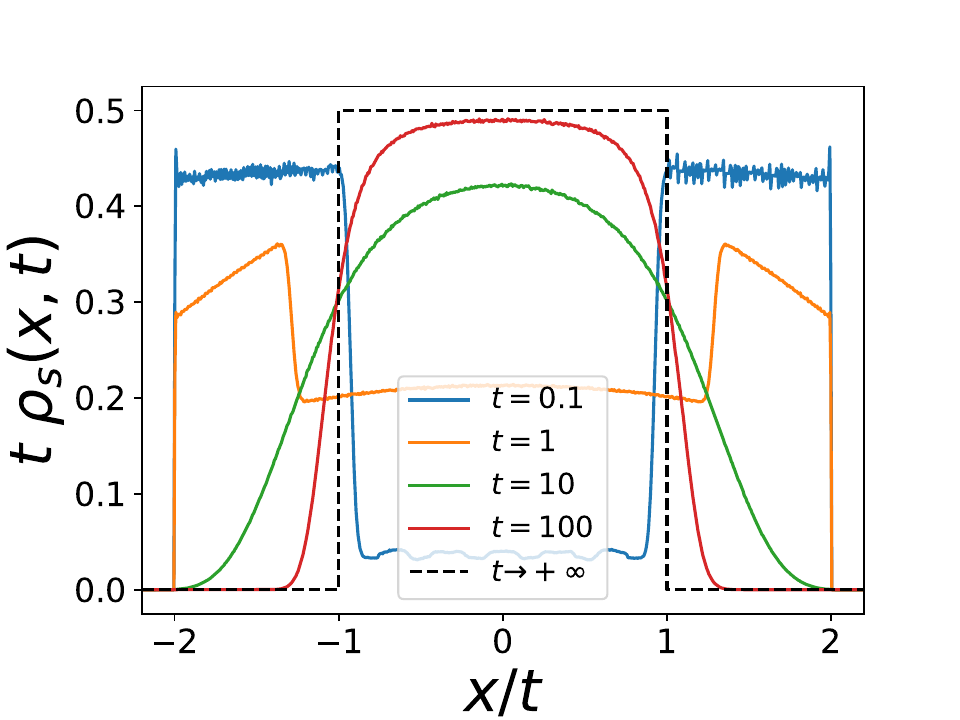}
    \includegraphics[width=0.49\linewidth]{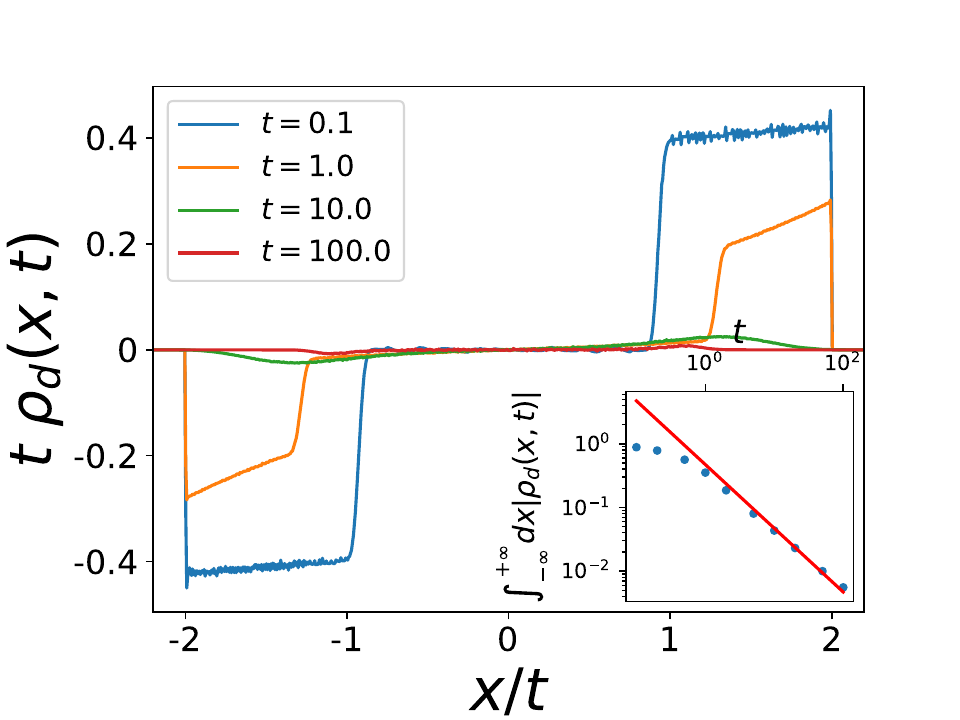}
    \includegraphics[width=0.49\linewidth]{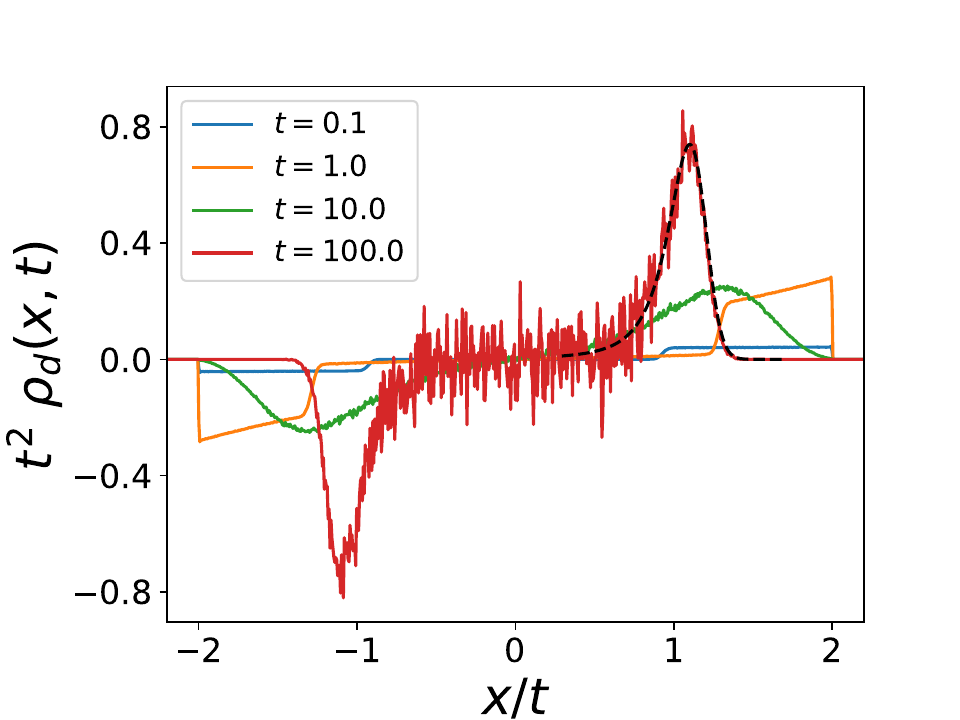}
    \includegraphics[width=0.49\linewidth]{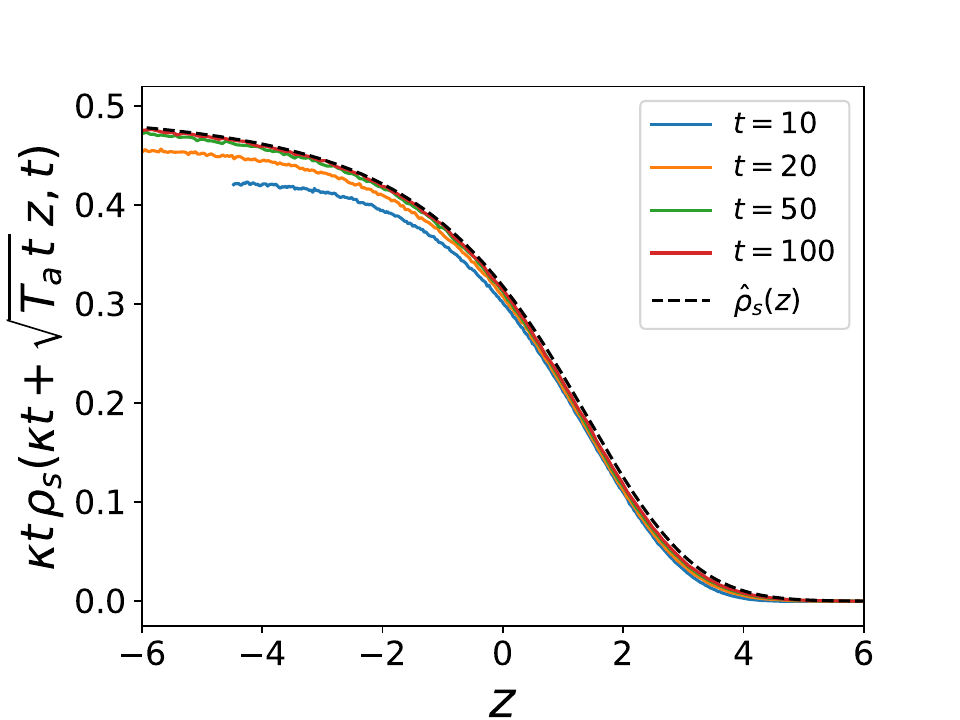}
    \caption{{\bf Top Left:} Density $\rho_s(x,t)$, obtained through simulations, plotted at different times as a function of the scaled position $x/t$, for the repulsive gas with $v_0=1$, $\kappa=1$ and $\gamma=1$, for $N=1000$ particles. At $t=0$ all particles are at $x=0$. The dashed black curve shows the analytical prediction for $t \to +\infty$. {\bf Top Right:} Same plot for $\rho_d(x,t)$. The inset shows the integral of the absolute value as a function of time. It is in good agreement with a $1/t$ decay (red line). {\bf Bottom left:} Plot of $\rho_d(x,t)$ multiplied by $t^2$ to compensate for the $1/t$ decay. The dashed black line shows the analytical prediction for the boundary layer \eqref{BL} for $t=100$. {\bf Bottom right:} Density $\rho_s(x,t)$ near the edge of the plateau, plotted at different times with the boundary layer scaling \eqref{BL}. At large times the scaled density converges to $\hat \rho_s(z)$.
    }
    \label{fig_repulsive}
\end{figure}

{\bf Repulsive case.} We now consider the repulsive case $\kappa=-\bar \kappa>0$, still for $V(x)=0$,
\bea \label{system1rep} 
&& \partial_t r = T \partial_x^2 r - v_0 \partial_x s  - 2 \kappa \, r \partial_x r \;, \\
&& \partial_t s = T \partial_x^2 s - v_0 \partial_x r - 2 \kappa r \partial_x s - 2 \gamma s \;. \label{system2rep} 
\eea
Since there is no confining potential to balance the repulsion between particles, all the particles escape at infinity and there is no stationary state. 
As in the case of passive particles, studied in \cite{PLDRankedDiffusion, FlackRD}, one expects that the size of the cloud grows linearly with time.
Hence we will look for a large time solution of (\ref{system1rep}-\ref{system2rep}), as a scaling function of $y=x/t$ 
in the form
\be \label{expansion}
r(x,t)= r_0(y)  + \frac{r_1(y)}{t} + \dots \ , \ 
s(x,t) = \frac{s_1(y)}{t}  + \dots \ , \ y=\frac{x}{t} 
\ee 
At leading order 0 one has
\be 
y r_0'(y) = 2 \kappa \, r_0(y) r_0'(y) \;.
\ee 
The solution is either $r_0'(y)=0$ or $r_0(y)=y/(2k)$ i.e. one has
\be 
r_0(y) = \begin{cases} -\frac{1}{2} \quad {\rm for} \ y < -\kappa \\ 
\frac{y}{2 \kappa} \quad {\rm for} \ -\kappa < y < \kappa \\
\frac{1}{2} \quad {\rm for} \ y > \kappa
\end{cases}
\ee
which leads to an expanding plateau
\be \label{rhos_repulsive}
\rho_s(x,t) = \partial_x r(x,t) \simeq \frac{1}{2 \kappa t} \, \theta(\kappa t - |x|) 
\ee 
where $\theta(x)$ is the Heaviside distribution. Hence the scaled density $\rho_s$ takes the form of an expanding plateau, exactly as for passive particles
\cite{PLDRankedDiffusion, FlackRD}. This is not too surprising since at large time the particles
are far from each other hence the active noise averages out to diffusive noise. There are 
however non trivial fluctuations with slow decay at large time. Indeed, the 
second equation gives
\be \label{repulsive_sr_rel}
s_1(y) =  - \frac{v_0}{2 \gamma}  r_0'(y) = - \frac{v_0}{4 \kappa \gamma} \, \theta(\kappa-|y|) 
\ee 
hence one finds 
\bea
&& s(x,t) \simeq - \frac{v_0}{4 \kappa \gamma t} \, \theta(\kappa t - |x|) = - \frac{v_0}{2 \gamma} \partial_x r(x,t) \;, \\
&& \rho_d(x,t) \simeq - \frac{v_0}{4 \kappa \gamma t} (  \delta(x+\kappa t) - \delta(x-\kappa t) ) \;. \label{rhod_repulsive}
\eea
Thus $\rho_d=0$ inside the plateau, i.e. $\rho_+=\rho_-$ (to leading order in $1/t$) 
but there is an excess of $-$ particles on the left edge and $+$ on the right edge
with total weight decaying as $1/t$. 

It is important to note that the $1/t$ expansion \eqref{expansion} is valid inside the
plateau \cite{footnoter1} but fails at the edges. A more careful analysis 
in a region of width $\sim \sqrt{t}$ near the edges shows \cite{SM} that
the solution takes a boundary layer form (at the right edge)
\bea \label{BL} 
\!\!\!\!\!\!\!\!\!\!\!\! &&\rho_s(x,t) = \frac{1}{\kappa t} \hat \rho_s (z)  \ , \ \rho_d(x,t) = -\frac{v_0}{2 \gamma \kappa \sqrt{T_{\rm eff}} \, t^{3/2}} \hat \rho_s' (z) \nonumber \\
\!\!\!\!\!\!\!\!\!\!\!\! &&\hat \rho_s(z) = \frac{e^{-\frac{z^2}{2}}(2+\sqrt{\pi} e^{\frac{z^2}{4}} z \, {\rm erfc}(-\frac{z}{2}))}{2\pi \, {\rm erfc}(-\frac{z}{2})^2}  \ , \ z =\frac{x-\kappa t}{\sqrt{T_{\rm eff} t}}
\eea
with $T_{\rm eff} = T + T_a = T + \frac{v_0^2}{2 \gamma}$. The boundary layer scaling
function $\hat \rho_s(y)$ is identical to the one obtained for the passive problem 
\cite{FlackRD}. This shows that for the expanding repulsive gas the role of the active noise is not different from that of passive noise 
at temperature $T_a=\frac{v_0^2}{2\gamma}$. Accordingly, note that the relation $\rho_d \simeq - \frac{v_0}{2 \gamma} \partial_x \rho_s$
holds at large time both in the boundary layer and in the plateau (which we have checked numerically). These results show that the delta function in \eqref{rhod_repulsive} has a finite width at finite time, i.e contrary to the attractive case there are no clusters.
In fact the $\pm$ population imbalance $\sim 1/t$ is subdominant compared to 
the total population in the boundary layer $\sim 1/\sqrt{t}$. 

We have checked these predictions numerically. Figure \ref{fig_repulsive} shows the convergence of the density $\rho_s$ to 
\eqref{rhos_repulsive} as time increases. The signature of the active noise is visible at short times: for $t < 1/\gamma$, the particles split into two packets with a spread in velocities $\frac{x}{t} \in [v_0,v_0+ \kappa]$ (and same with minus), leading to two distinct blocks in the density, 
which disappear at larger time as the active noise averages out. One also sees in Figure \ref{fig_repulsive} peaks
forming the scaled $t^2 \rho_d(x, t)$
as well as a numerical check of the boundary layer forms \eqref{BL}. 



{\bf Linear potential.} We now consider the case where the particles are submitted to
an attractive linear potential $V(x)=a|x|$, $a>0$. For non-interacting particles
$\bar\kappa=0$ there is a steady state with two phases, either $a<v_0$ and 
\be \label{potential_N1}
\rho_s(x) = \frac{\gamma a}{v_0^2-a^2} \, \exp \left(-\frac{2\gamma a}{v_0^2-a^2}|x| \right)
\ee
or $a>v_0$ and $\rho_s(x)=\delta(x)$ \cite{DKM19}. 

\begin{figure}
    \centering
    \includegraphics[width=0.95\linewidth, trim={0.8cm 3.1cm 0.8cm 4.7cm}, clip]{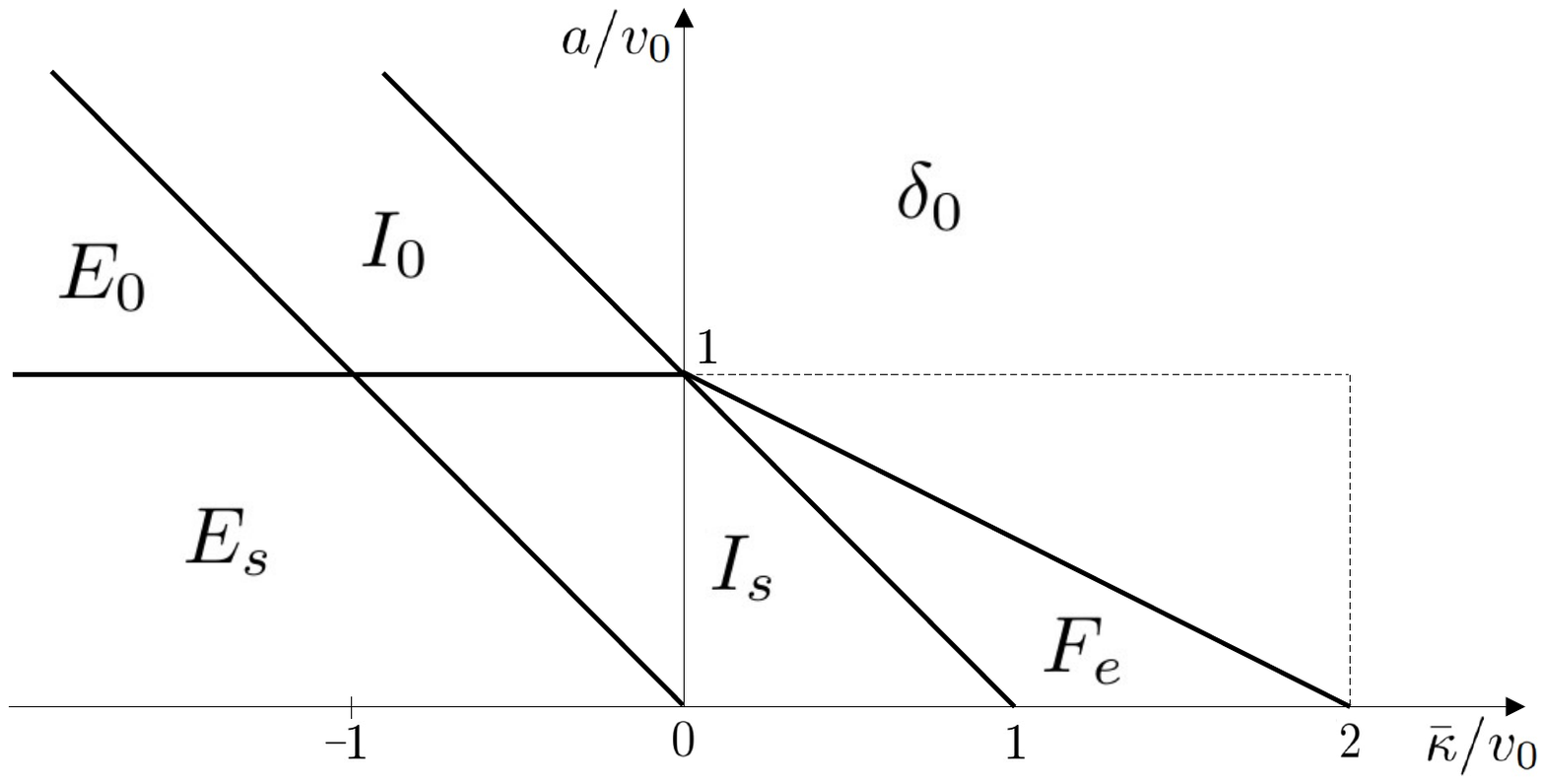}
    \caption{Phase diagram of active ranked diffusion with a linear external potential $V(x)=a|x|$. The phases are named as follows: the large letter is either $E$ for an expanding phase, $I$ for a stationary phase with infinite support, $F$ if the support is finite support, or $\delta$ for a clustered phase. The subscript indicates the position of the shocks: $0$ if there is a shock at $x=0$, $e$ if there are shocks at the edges and $s$ if the density is smooth. The dotted lines denote a finite $N$ transition.}
    \label{phase_diagram_potential}
\end{figure}

In presence of interactions, the large time behavior for large $N$ is
richer and represented on the phase diagram in Fig. \ref{phase_diagram_potential}. 
There are six distinct phases which we briefly describe, four of them being stationary, while the other two are expanding (see \cite{SM} for details and some plots of the densities).

{\it $\delta_0$ phase}. It exists both in the attractive case, for $a>v_0 - \frac{\bar \kappa}{2}$,
and in the repulsive case, $a> v_0 + \kappa$. The density is formed of a single cluster at $x=0$, $\rho_s(x)=\delta(x)$.
Note that outside the dotted lines in Fig. \ref{phase_diagram_potential} it already exists for finite $N$,
while inside the density has a finite width for finite $N$. 

{\it $I_s$ phase,} both in the attractive case, for $a<v_0 - \bar \kappa$, and in the repulsive case, for $\kappa < a < v_0$. 
This phase is quite similar to the one for $a=0$. The function $r=r(x)$ is odd and 
is given for $x \geq 0$ as the solution of
\be \label{eqres_potential0}
\gamma x = {\sf f}_a(r)= 2 \bar \kappa r + \frac{v_0^2 - (\bar \kappa+a )^2 }{2(\bar \kappa+a)} \log\left( \frac{\bar \kappa + 2 a + 2 \bar \kappa r}{(\bar \kappa + 2 a)(1-2 r) } \right) 
\ee
an extension of \eqref{solu0} above. From $r(x)$ one obtains for any $x \in \mathbb{R}$
\be \label{s_function_r_lin0}
s(x) = \frac{\bar \kappa}{v_0} \big(r(x)^2 - \frac{1}{4} \big) + \frac{a}{v_0} \big(|r(x)| - \frac{1}{2} \big) \;.
\ee 
The densities are given in parametric form for $x \geq 0$ and $r \in [0,1/2]$ as
\be \label{rho_parametric}
x= \frac{{\sf f}_a(r)}{\gamma} ~,~ \rho_s = \frac{\gamma}{{\sf f}'_a(r)} 
~,~ \rho_d = \frac{(a + 2 \bar \kappa r)\gamma}{v_0 {\sf f}'_a(r)} \;.
\ee 
They have infinite support and are smooth except at $x=0$ where $\rho_d(x)$ has a jump and 
$\rho_s(x)$ has a linear cusp. 

{\it $F_e$ phase,} in the attractive case, for $v_0 - \bar \kappa < a < v_0 - \frac{\bar \kappa}{2}$. 
Here ${\sf f}_a(r)$ becomes non-invertible. The densities have a finite support $[-x_e,x_e]$
with shocks at the two edges $\pm x_e$ such that
\be \label{rxe} 
r(x_e^-) = \frac{v_0-a}{\bar \kappa} - \frac{1}{2}  \quad , \quad r(x_e^+) = \frac{1}{2}
\ee
with $\gamma x_e= {\sf f}_a(r(x_e^-))$. It corresponds to clusters with a fraction $\frac{\bar \kappa + a - v_0}{\bar \kappa}$ of the particles
(all being $+$ at $x=x_e$ and $-$ at $x=-x_e$). 
Inside the support, $r(x)$ is again given by \eqref{eqres_potential0},
$s(x)$ by \eqref{s_function_r_lin0} and the densities by \eqref{rho_parametric},
with the same singular behavior at $x=0$.

{\it $I_0$ phase,} in the repulsive case, for $\kappa<a<v_0+ \kappa$ and $a>v_0$.
This phase is similar to $I_s$ but has a shock at $x=0$ corresponding to
a cluster with equal fractions of $\pm$ particles and of total weight 
$\frac{a-v_0}{\kappa}$. The function $r(x)$ for $x>0$ is now given by
\be
\gamma x = {\sf f}_a(r) - {\sf f}_a(r(0^+)) ~,~ r(0^+) = \frac{a-v_0}{2\kappa} \;.\label{rI0} 
\ee
From \eqref{rI0}, \eqref{eqres_potential0} and \eqref{s_function_r_lin0} (which still holds 
for $x \neq 0$) we obtain that $\rho_s(x)$ and $\rho_d(x)$ exhibit near $x=0$ an inverse square root divergence 
\be
 \rho_s(x) \simeq \frac{a-v_0}{\kappa}  \delta(x) + \frac{B}{2 \sqrt{|x|}} \ , \ \rho_d(x) \simeq  \frac{B}{2 \sqrt{|x|}} \, {\rm sgn}(x) 
\ee
with $B= \frac{1}{2\kappa} \sqrt{\frac{\gamma}{v_0} (v_0^2 - (a- \kappa)^2)}$, which
involves only $+$ particles for $x>0$ and
$-$ particles for $x<0$. 

{\it $E_s$ and $E_0$ phases,} in the repulsive case, for $a<\kappa$ and $a<v_0$ or $a>v_0$ respectively.
In this case a fraction $1-\frac{a}{\kappa}$ of particles escape to infinity,
while a fraction $\frac{a}{\kappa}$ remain confined. Under the scaling
$|x| \sim t$ one finds at large time in both phases
\be
\rho_s(x,t) \simeq \frac{a}{\kappa} \delta(x) + \frac{1}{2\kappa t} \theta((\kappa-a)t - |x|) \;, 
\ee
hence the expanding part behaves similarly as for $a=0$, i.e. it is uniform within
the support, which now spreads
with velocities $\pm (\kappa-a)$. Similarly one finds
\be
\rho_d(x,t) \simeq \frac{v_0}{4\gamma\kappa t} (\delta((\kappa-a)t - x) - \delta((\kappa-a)t - x)) \;.
\ee
Both densities are smooth on scales $\sqrt{t}$, with similar boundary layers 
near the edges as for $a=0$ (see above and \cite{SM}).

For the particles which remain bound within $x=O(1)$, $r(x)$ is smooth away from $x=0$ given by, for $x>0$
\bea \label{rE}
&& \!\!\!\!\!\! \gamma x = \tilde {\sf f}_a\left(\frac{\kappa}{a} r(x)\right) - \tilde {\sf f}_a\left(\frac{\kappa}{a} r(0^+)\right) \;, \\
&& \!\!\!\!\!\! \tilde {\sf f}_a(r)= 2 a r ( \frac{v_0^2}{a^2(1-2r)} - 1 ) \ , \ r(0^+) = \frac{a-v_0}{2\kappa} \theta(a-v_0) \nonumber 
\eea
where in the phase $E_0$, i.e. $a>v_0$, there is a cluster at $x=0$ containing a fraction $\frac{a-v_0}{\kappa}$
of the total number of particles. From \eqref{rE} one obtains an explicit formula for the densities, see \cite{SM}.
The most notable feature is that they decay {\it as power laws} at large distance, in both phases
\be 
\rho_s(x) \simeq \frac{v_0^2}{2 \kappa \gamma x^2} \quad , \quad \rho_d(x) \simeq \frac{v_0^3}{2 \kappa \gamma^2 x^3} \;,
\ee
while in the phases $I_s$ and $I_0$ the densities decay exponentially at large distance.
In that sense in the phases $E_s$ and $E_0$ the steady state of the bound particles is always critical. Indeed, the bound particles 
can be effectively described by an interaction constant $\kappa_{\rm eff}=a$, see \cite{SM}. Finally,
the non-analytic behavior of the densities near $x=0$ is similar to the
one in the phases $I_s$ and $I_0$. In \cite{SM} we compare these results with the case of passive particles, where only the phases $I_s$ and $E_s$ are present.

{\bf More general interactions}. Consider now a pairwise interaction potential $W(x)$,
such that the force term in \eqref{langevin1} is replaced by $- \frac{1}{N} \sum_j W'(x_i-x_j)$,
Eq. \eqref{langevin1} being recovered for $W(x)=\bar \kappa |x|$. $W'(x)$ being odd, we restrict to the 
case $W'(0)=0$, i.e. we exclude the possibility that $W'(x)$ diverges at $x=0$ for which the large $N$ 
limit is problematic \cite{TouzoDBM2023}. At large $N$, from \eqref{eqrho1} 
with the replacement $\bar \kappa \, {\rm sgn}(x-y) \to W'(x-y)$ we obtain at stationarity
\bea \label{eqrho2n_text}
&& 0 =    \partial_x [ (- v_0 \rho_d - \tilde F(x) \rho_s ] \\
&& 0 =   \partial_x [ (- v_0 \rho_s - \tilde F(x) \rho_d ] - 2 \gamma \rho_d\nn
\eea 
where
\be \label{def_tildeF}
\tilde F(x) = - V'(x) - \int dy \, W'(x-y) \rho_s(y) \;.
\ee 
Eqs. \eqref{eqrho2n_text} are identical to those of a single RTP in an effective force field $\tilde F(x)$
which itself depends on the density. They can be solved formally, leading (in the simplest case, see \cite{SM}) 
to 
\be 
\rho_s(x) = \frac{K}{v_0^2-\tilde F(x)^2} e^{2 \gamma \int_0^x dz \frac{\tilde F(z)}{v_0^2-\tilde F(z)^2}} \label{self_text}
\ee 
where $K$ is determined by the normalization. Eqs. \eqref{def_tildeF} and \eqref{self_text} are self-consistent equations for the density $\rho_s(x)$.
In \cite{SM} we show how it recovers the above results for the rank interaction. 
For a harmonic interaction the density at large $N$ is found to be the same as for a single particle
in a harmonic well. Other examples will be studied in
\cite{inprep}. 

{\bf Conclusion}. We have obtained a description of an active version
of rank diffusion, controlled at large $N$. We have computed exactly the
densities of RTP particles at large time in that limit. In particular,
in the attractive case we found a collective dynamical phase transition in the steady state
where the support of the density changes from infinite to finite, with shocks (i.e
clusters of particles) appearing at the edges. In presence of a linear confining
potential we find a rich phase diagram with six phases. In the expanding
phases we find that a fraction of the particles remain bound, and in
a critical state with power-law decay of the densities at large distance.
We also obtained a self-consistent equation for the densities which opens the way to treat
more general interaction potentials. Some remaining open questions are
finite $N$ fluctuations and calculation of other observables, such as entropy production.

\bigskip

{\it Acknowledgments:} 
We thank Gregory Schehr for collaborations on related topics. 
We thank LPTMS (Orsay) and Coll{\`e}ge de France for hospitality.

{}

\newpage

.

\newpage

.

\newpage

\begin{widetext}

\setcounter{secnumdepth}{2}

\begin{large}
\begin{center}

Supplementary Material for\\  {\it Active rank diffusion}

\end{center}
\end{large}

\tableofcontents

\bigskip

We give the principal details of the calculations described in the main text of the Letter. 
We display additional analytical and numerical results which support the findings presented in the Letter. 

\bigskip


\section{A few definitions and properties}

\subsection{Center of mass and density fields for finite $N$}

The center of mass for any $N$ is defined by
\be 
\bar x(t)=  \frac{1}{N} \sum_i x_i(t) \;.
\ee 
Its equation of motion is given by 
\be 
\frac{d}{dt} \bar x = \frac{v_0}{N} \sum_i \sigma_i(t) + \sqrt{2T} \xi_i(t)
\ee 
i.e. it evolves as the sum of the positions of $N$ independent RTP's with velocities $v_0/N$.
Since $\langle \sigma_i(t) \sigma_j(t') \rangle = \delta_{ij} e^{-2 \gamma |t-t'|}$, one
obtains that for $\gamma>0$ at large times the center of mass has a diffusive behavior $\bar x \sim \sqrt{2 D_N t}$, 
with 
\be 
D_N=\frac{1}{N}(T + \frac{v_0^2}{2 \gamma}) \;.
\ee

At finite $N$ one can define two density fields 
\bea
&& \rho_{\sigma}(x,t) = \frac{1}{N} \sum_i \delta_{\sigma,\sigma_i(t)} \delta(x-x_i(t)) \\
&& \hat \rho_{\sigma}(x,t) = \frac{1}{N} \sum_i \delta_{\sigma,\sigma_i(t)} \delta(x-x_i(t)- \bar x(t)  ) \quad , \quad \bar x(t) = \frac{1}{N} \sum_j x_j(t) \;.
\eea 
The first definition is the one used in the main text, for which we derive a Dean equation (valid in principle for any $N$, but
studied mostly for large $N$). The second definition is the one we use for the numerical simulations (except when a confining
potential is present, in which case we use the first definition). The difference between the
two definitions becomes irrelevant in the large $N$ limit since $D_N \to 0$ in that limit. We will return to
this point below. 

\subsection{Results for $N=2$} \label{app:N2}

As an example let us recall the explicit result for the stationary distribution for $N=2$ 
obtained in \cite{us_bound_state} in the attractive case $\bar \kappa>0$. 
Here we consider the variable $z=\frac{x_1-x_2}{2}$, while there we denoted $y=x_1-x_2=2 z$.
We also have the correspondence $\bar \kappa = 2 \bar c$ with the notations of 
\cite{us_bound_state}). Taking this into account, the stationary probabilities
$P_{\sigma_1,\sigma_2}(z)$ for the variable $z$ read,
for $v_0 > \bar \kappa/2$,
\bea \label{result2} 
 \left(
\begin{array}{c}
 P_{++}(z)  \\
P_{+-}(z)  \\
P_{-+}(z)  \\
P_{--}(z)   \\
\end{array} \right) 
 =
 \frac{\gamma \bar \kappa}{4v_0^2 + \bar\kappa^2}\;  
 e^{- \frac{8 \gamma \bar \kappa}{4 v_0^2 - \bar \kappa ^2} |z|} 
 \left(
\begin{array}{c}
 1  \\
\frac{2 v_0+\bar \kappa}{2v_0-\bar \kappa}  \theta(z) + 
\frac{2v_0-\bar \kappa}{2v_0+\bar \kappa}  \theta(-z) \\
\frac{2v_0-\bar \kappa}{2v_0+\bar \kappa}   \theta(z) + \frac{2v_0+\bar \kappa}{2v_0-\bar \kappa}  \theta(-z)  \\
1  \\
\end{array}
\right)  + \frac{1}{2} \frac{\bar \kappa^2}{4 v_0^2 + \bar \kappa^2}
 \delta(z) \left(
\begin{array}{c}
 1 \\
0 \\
0 \\
1  \\
\end{array} \right)  \;.
\eea
The total probability is $P(z)=\sum_{\sigma_1=\pm 1,\sigma_2=\pm 1} 
P_{\sigma_1,\sigma_2}(z)$, which is normalized to $\int_{-\infty}^{+\infty} dy P(y)=1$.
One can check from \eqref{result2} 
that one recovers the expression of the $P(z)$ given in the text in 
\eqref{tot_prob1}. Note the symmetry $P_{\sigma_1,\sigma_2}(z)=P_{\sigma_2,\sigma_1}(-z)$
and that for each state $(\sigma_1,\sigma_2)$ one has
$\int_{-\infty}^{+\infty} dy P_{\sigma_1,\sigma_2}(y) = \frac{1}{4}$, i.e. each of the four states is equiprobable in the steady state,
as expected. 

From this one can compute the second definition of the density, i.e. the
density in the reference frame of the center of mass. It is given by 
\bea
&& \hat \rho_\sigma(x) = \frac{1}{2} \langle \delta_{\sigma_1,\sigma} \delta(x_1-\bar x - x) \rangle + 
\frac{1}{2} \langle \delta_{\sigma_2,\sigma} \delta(x_2- \bar x -x) \rangle \\
&& = \frac{1}{2} \langle \delta_{\sigma_1,\sigma} \delta(z - x) \rangle 
+ \frac{1}{2} \langle \delta_{\sigma_2,\sigma} \delta(-z -x) \rangle \\
&& = \frac{1}{2} \sum_{\sigma'} P_{\sigma,\sigma'}(z) + 
\frac{1}{2} \sum_{\sigma'} P_{\sigma',\sigma}(-z) = \sum_{\sigma'} P_{\sigma,\sigma'}(z) 
\eea 
where we used the above mentioned symmetry. Hence we obtain
\bea 
\hat \rho_s(x) = \hat \rho_+(x) + \hat \rho_-(x) = P(x) 
\eea 
where $P(x)$ is the total probability given in the text in \eqref{tot_prob1},
and
\bea 
\hat \rho_d(x) = \hat \rho_+(x) - \hat \rho_-(x) = P_{++}(x) + P_{+-}(x) - P_{-+}(x) - P_{--}(x) =  {\rm sgn}(x)  \frac{8 \gamma v_0 \bar \kappa^2}{16 \, v_0^4 - \bar \kappa^4}\;  
 e^{- \frac{8 \gamma \bar \kappa}{4 v_0^2 - \bar \kappa^2} |x|}  \,.
\eea 
Note that for $v_0< \bar \kappa/2$ one finds \cite{us_bound_state} that $P_{++}(z)=P_{--}(z)=\frac{1}{2} \delta(z)$ and
$P_{+-}(z)=P_{-+}(z)=0$. Hence the density $\hat \rho_s(x)=\delta(x)$, $\hat \rho_d(x)=0$, 
and the two particles form a single cluster. 


\subsection{Center of mass and median point at large $N$}

In terms of the rank field $r(x,t)$ the position of the center of mass is given for any $N$ as
\be 
\bar x(t)=  \frac{1}{N} \sum_i x_i(t)= \int_{-\infty}^{+\infty} dx \, x \partial_x r(x,t) = \int_{-1/2}^{+1/2} dr \, x(r,t) \;.
\ee 

Consider now the limit $N \to + \infty$. In that limit, the fractions $p_\pm(t)=N_\pm(t)/N$ of $\pm$ particles
follow a deterministic evolution equation, $\partial_t (p_+-p_-)=- 2 \gamma (p_+-p_-)$. Hence, for $\gamma>0$, 
$p_+(t)-p_-(t)$ decreases exponentially to zero at large $t$. 

Consider now the equations (\ref{system1}-\ref{system2}) valid for $N=+\infty$. Recall that $s(+\infty,t)=p_+(t)-p_-(t)$. 
In the case where 
the initial condition satisfies 
$p_+(0)=p_-(0)$, which implies $s(+\infty,t)=0$ for all $t$, one can check from these equations that
the center of mass does not move.
Indeed one obtains
\be 
\frac{d}{dt} \bar x(t) = \int_{-\infty}^{+\infty} dx \, x \partial^2_x ( T \partial_x r(x,t) - v_0 s(x,t) + \bar \kappa(r(x,t)^2 - \frac{1}{4}) ) = 0
\ee 
which vanishes by integration by part since the term inside the second derivative vanishes at $x=\pm \infty$. 

One can also define the median point $x_0(t)$, which for any $N$ odd, and for $N=+\infty$ is such that
\be
x_0(t) = 
r(x_0(t),t) = 0 \quad , \quad x_0 = x(0,t) \;.
\ee 
This point however may move (depending on the initial condition) if the system has not reached the stationary state
(even if $p_+=p_-$).

\section{Numerical simulation details}
The simulations were performed using simple Euler dynamics. To make the computations faster, we use the fact that, as the name suggests, the total force applied on a particle due to the rank interaction only depends on the rank of the particle's position. We thus keep in memory the updated rank of each particle if they were ordered by position (taking into account the fact that some can have the same position if a cluster forms). The positions of all particles are then updated simultaneously at each time step according to
\be
x_i(t+dt) = x_i(t) + dt \left(\frac{\bar \kappa}{N}(N_i^{\rm right}(t)-N_i^{\rm left}(t)) -V'(x_i(t))+v_0 \sigma_i(t) \right) 
\ee
where $N_i^{\rm right}(t)$ (resp. $N_i^{\rm left}(t)$) is the number of particles strictly at the right (resp. left) of particle $i$ at time $t$, and the $\sigma_i$'s switch sign independently with probability $\gamma dt$ at each time step. Strictly speaking, one should keep track of every particles that cross and check if they form a new cluster. In practice, in the phases where no shocks are present in the density at finite $N$ (i.e. $I_s$, $F_e$, $E_s$ and $\delta_0$ inside the dotted rectangle), one can simply let the particles cross, in which case they start to perform very small oscillations around each other (if the time step is small enough for the bin size of the histogram, this situation can not be distinguished from a real cluster). In the presence of a linear potential however, we do keep track of clusters forming at $x=0$ (in the phases $I_0$ and $E_0$, i.e. whenever $a>v_0$). More precisely when a particle should cross $x=0$, we place it exactly at $x=0$ and wait for the next time step to see if it keeps moving. This avoids having a large number of particles oscillating around $x=0$, which would slow down the simulations.

For the phases where a steady state exists (or when looking at the bound particles in an expanding phase), the plots are obtained using a single run of the dynamics. We run the dynamics long enough to let the system reach stationarity, and then build the histogram of positions (after subtracting the position of the center of mass if no confining potential is present) over a large time window. In this case we use a time step $dt=0.001$ and run the dynamics for $10^7$ to $10^9$ steps, depending on the value of $N$. To look at the escaping particles in the expanding phases, we instead run the dynamics $10^5$ times and build a distinct histogram for each time step. In this case we use $dt=0.01$.

\section{More details about the particle clusters}
Let us first recall what the phase diagram in Fig. \ref{phase_diagram_potential} looks like at finite $N$. In the repulsive case it is not qualitatively different from the infinite $N$ case. In particular, the shock at $x=0$ for $a>v_0$ and the transition to an expanding crystal for $\kappa>a$ are effects that already exist at finite $N$ and are independent of $N$. In the attractive case the situation is a bit different. Indeed, in the range of parameters corresponding to the phase $F_e$, the delta peaks in the densities at $x=\pm x_e$ are replaced by peaks of finite width. Thus the transition from a smooth density to the presence of shocks is an effect of large $N$. Additionally, in the phase $\delta_0$, the density $\rho_s$ is a delta function even at finite $N$ only outside the dotted rectangle ($a>v_0$ or $\bar \kappa>2 v_0$) while inside it has a finite width which decreases with $N$. Below we will nevertheless use the names of the infinite $N$ phases, even at finite $N$, to refer to the range of parameters corresponding to these phases.


As mentioned in the text, for any finite $N \geq 2$ the particles tend to aggregate into clusters (i.e. groups of particles having all the same position for a finite period of time) due to the discontinuity in the interaction force $-\frac{\bar \kappa}{N} {\rm sgn}(x_j-x_i)$ (in the attractive case), as well as in the external force $-V'(x_i)=-a \, {\rm sgn}(x_i)$ when it is present \footnote{Note that if instead these discontinuities where smooth with a microscopic scale $l$, the clusters would still be present but would have a finite size $O(l)$.}. 
In the phases where the density is smooth, such clusters can only form if two or more particles keep the same value of $\sigma$ over an extended period of time, thus they are generally small. {For instance in the case $N=2$ without external potential, for $v_0>\bar \kappa/2$, which we recalled in App. \ref{app:N2}, this corresponds to the delta part in $P_{++}(z)$ and $P_{--}(z)$ in \eqref{result2}.} More precisely, we expect that the probability to observe a cluster containing any finite fraction of particles decays exponentially with $N$ in this case. 
Thus such clusters are not visible in the densities for $N \to +\infty$. The delta peak in the density $\rho_s$ at $x=0$ for finite $N$, with weight $c_N$ (see Fig. \ref{fig_smallN}, {which indeed shows the exponential decay in $N$}), corresponds to the particular case where all the particles form a single cluster, {an event occuring with probability $c_N$ in the steady state. Note that for finite $N$, only the clusters which have a fixed position (in the reference frame of the center of mass for the case $a=0$) are visible as delta peaks in the density. For $N\geq 3$ clusters of all sizes are present, but only clusters of size $N$ have a fixed position and thus appear in the density.}

However, larger clusters tend to form when the attraction between particles or the external potential become too strong compared to the noise. In particular, in the phase $F_e$, since a single particle at the edge is always attracted towards the center of mass, clusters containing a finite fraction of particles form near $x=\pm x_e$ (with a high probability which goes to $1$ as $N$ goes to infinity). However, the position of these clusters as well as the number of particles that they contain fluctuate. Thus they appear as peaks in the densities, with a finite width which decreases with $N$ (see right panel of Fig. \ref{fig_attractive}). These peaks only become delta functions in the limit $N \to +\infty$. This is also the case for the phase $\delta_0$ inside the dotted rectangle in Fig. \ref{phase_diagram_potential}.

Finally, in the phases $I_0$, $E_0$ and $\delta_0$ (outside the dotted rectangle), we observe the formation of stable clusters, which have a fixed position, in that case $x=0$ (in the reference frame of the center of mass for the case $a=0$), 
and contain a fixed number of particles, corresponding to a finite fraction of the particles, independent of $N$. In this case, this leads to a delta peak in the density even at finite $N$, with a weight which is independent of $N$.

All these statements could be made more quantitative through a numerical study of the distribution of cluster sizes in the different phases (and its dependence on the number of particles $N$). This analysis goes beyond the scope of the present paper and is left for future work.

\section{Derivation of the main equations}

Using the Dean-Kawasaki approach  \cite{Dean,Kawa}, one can establish, as in \cite{PLDRankedDiffusion} for the passive case,
and as in Sec. III A of the Supp Mat. in \cite{TouzoDBM2023} in the active case
(replacing in (74) arXiv version $W'(x) \to V'(x)$, $\tilde V'(x) \to \frac{\bar \kappa}{N} {\rm sgn}(x)$, $\tilde \rho \to \rho_+ + \rho_-$, 
and $T \to N T$), a stochastic evolution equation
for the density fields, which takes the following form
\bea \label{eqrho_full}
&& \partial_t \rho_\sigma(x,t)  =  T \partial_x^2 \rho_\sigma(x,t)  +  \partial_x [\rho_\sigma(x,t)  ( - v_0 \sigma +  V'(x) + \bar \kappa \int dy (\rho_+(y,t) + \rho_-(y,t) ) {\rm sgn}(x-y) )] \\
&& + \gamma \rho_{- \sigma}(x,t) - \gamma \rho_{\sigma}(x,t) 
+ \frac{1}{\sqrt{N}} \partial_x \xi_{\sigma}(x,t) + \frac{\sigma}{\sqrt{N}}\zeta(x,t) \;. \nn
\eea
Here $\xi_\pm(x,t)= \sqrt{2 T \rho_\sigma(x,t)} \eta_\sigma(x,t)$ are two independent demographic passive noises,
where $\eta_\pm$ are two independent standard space time white noises, 
originating from the thermal white noises, and $\zeta(x,t)$ is a non-Gaussian active noise, white in time,
originating from the telegraphic noises. Some more details about these noises are provided in Sec. III A of the Supp Mat. in \cite{TouzoDBM2023}.

It is very important to note that we have used the fact that the interaction force vanishes at $x=0$, i.e. ${\rm sgn}(0)=0$, in the present model, 
which allows to establish the DK equation without difficulties (see the discussion in Sections III A and III B of the Supp Mat. in \cite{TouzoDBM2023}).
Physically, this is related to the fact that here the particles can cross. 

Equation \eqref{eqrho_full} is formally exact for any $N$. However it is most useful at large $N$. 
Indeed both noise terms are typically $O(1/\sqrt{N})$, hence for most applications one can either neglect 
them or treat them in perturbation. At this stage the interaction term is still in a non-local form. However,
for the Coulomb interaction, it can be brought to a local form, as used in \cite{PLDRankedDiffusion},
and as we now discuss. Another case of long range interactions
where this is possible (and maybe the only other one)
is the logarithmic interaction, e.g. for the Dyson Brownian motion, where the
resolvant also obeys, in some cases, a local equation (see e.g. \cite{TouzoDBM2023}).

Recalling the definitions $\rho_s(x,t)=\rho_+(y,t) + \rho_-(y,t)$ and $\rho_d(x,t) = \rho_+(y,t) - \rho_-(y,t)$, we now study the large $N$ limit. Neglecting the noise (hence assuming that the density become self-averaging
in that limit) we obtain the pair of deterministic dynamical equations
\bea \label{eqrho2}
&& \partial_t \rho_s(x,t)  =  T \partial_x^2 \rho_s(x,t)  +  \partial_x [ (- v_0 \rho_d(x,t) + \rho_s(x,t) (V'(x) +
\bar \kappa \int dy \rho_s(y,t) {\rm sgn}(x-y) )] \;, \\
&& \partial_t \rho_d(x,t)  =  T \partial_x^2 \rho_d(x,t)  +  \partial_x [ (- v_0 \rho_s(x,t) + \rho_d(x,t) (V'(x) +
\bar \kappa \int dy \rho_s(y,t) {\rm sgn}(x-y) )] - 2 \gamma \rho_d(x,t) \;. \nn
\eea 
We will search for solutions of these equations with initial conditions 
$\rho_s(x,t=0)$, $\rho_d(x,t=0)$ vanishing for $x = \pm \infty$.

As in the main text, and in \cite{PLDRankedDiffusion} for the passive case, we now define 
{\it rank fields} $r(x,t)$ and $s(x,t)$ (also called height, or counting fields) through
\bea
&& \rho_s(x,t) = \partial_x r(x,t) \quad , \quad r(x,t) = \int^x_{-\infty} dx' \rho_s(x',t) - \frac{1}{2} \;, \\
&& \rho_d(x,t) = \partial_x s(x,t) \quad , \quad s(x,t) = \int_{-\infty}^x dy \, \rho_d(y,t) \nn \;.
\eea 
Hence $r(x,t)$ increases monotonically from $-1/2$ at $x=-\infty$ to $+1/2$ at $x=+\infty$,
while $s(x,t)$ vanishes for $x \to - \infty$. 
The equations \eqref{eqrho2} become
\bea
&& \partial_t \partial_x r(x,t) =  T \partial_x^3 r(x,t) + \partial_x[ - v_0 \partial_x s(x,t) 
+ 2  \bar \kappa r(x,t) \partial_x r(x,t)
+  V'(x) \partial_x r(x,t) ] \;, \\
&& \partial_t \partial_x s(x,t)  =  T  \partial_x^3 s(x,t)  +  \partial_x [ - v_0 \partial_x r(x,t)  + (V'(x) +
2 \bar \kappa  r(x,t)) \partial_x s(x,t) ] - 2 \gamma \partial_x s(x,t) \;, \nn
\eea 
where we have used that upon integration by part
\bea
\int dy \rho_s(y,t) {\rm sgn}(x-y) &=& \int dy \partial_y r(y,t) {\rm sgn}(x-y)
= 2 r(x,t) + [r(y,t) {\rm sgn}(x-y)]^{+\infty}_{-\infty} \nn \\
&=& 2 r(x,t) 
- (r(-\infty,t) + r(+\infty,t)) = 2 r(x,t) \;.
\eea

We can now integrate the first equation over $x$ noting that the integration constant vanishes from the boundary conditions for $r(x,t)$ at
$x=\pm \infty$ and since $\rho_s(x)$ and $\rho_d(x,t)$ vanish at infinity. We also make the natural assumption that $V'(x) \rho_s(x,t)$
vanishes at infinity. We can integrate also the second equation from $-\infty$ to $x$, which gives our
final set of dynamical equations
\bea \label{system}
&& \partial_t r(x,t) = T \partial_x^2 r(x,t) - v_0 \partial_x s(x,t)  + 2 \bar \kappa r(x,t) \partial_x r(x,t) 
+ V'(x) \partial_x r(x,t) \;, \\
&& \partial_t s(x,t)  =  T  \partial_x^2 s(x,t)  - v_0 \partial_x r(x,t)  + 2  \bar \kappa  r(x,t) \partial_x s(x,t)  + V'(x) \partial_x s(x,t) - 2 \gamma s(x,t) \;. \nn
\eea


These are the equations given in the text, and they are valid both for attractive and repulsive gas. 
Denoting $p_\pm(t)= N_\pm(t)/N$ the fraction of $\pm$ particles in the system at time $t$,
we note that with this choice of definition, $s(+\infty,t)= \int_{-\infty}^{+\infty} dx \rho_d(x,t) = p_+(t) - p_-(t)$ is not necessarily zero at finite time, depending on the initial condition. 
However, for $\gamma>0$ it decays to zero, since $\partial_t \int dx \rho_d(x,t) = - 2 \gamma \partial_t \int dx \rho_d(x,t)$. Hence for $\gamma>0$ (i) if one starts with $p_+(t=0)=p_-(t=0)$ then $p_+(t)-p_-(t)=0$ for all times and
(ii) in the stationary state, one has $p_+=p_-$. In both cases,
the field $s(x,t)=s(x)$ should vanish at both $x=\pm \infty$. The above properties are 
true of course only because here we have neglected the noise terms at large $N$.
Note that $\gamma=0$ is a special case, with different and conserved values of the ratios $p_\pm$,
where there can be many stationary states called fixed points.

{\bf Remark}. In the diffusive limit $v_0,\gamma \to +\infty$ with 
$T_a=\frac{v_0^2}{2 \gamma}$ fixed, one sees that it is natural to assume that
in the second equation in \eqref{system} all terms
but two are negligible and one gets 
\be 
- v_0 \partial_x r(x,t)   - 2 \gamma s(x,t) = 0 \;.
\ee 
Inserting in the first equation in \eqref{system} one sees that the term $- v_0 \partial_x s(x,t)$ 
tends to a thermal term $T_a \partial_x^2 r(x,t)$, and one recovers the equation for $r(x,t)$
of the passive case, with a thermal noise at temperature $T_a$.



\section{Large time solution in the presence of a linear potential}

In this section we consider the equations (\ref{system1}-\ref{system2}) which describe the system at large $N$ in the presence of a linear potential $V(x) = a |x|$ ($a \geq 0$), both in the attractive and repulsive case setting $T=0$.
This leads to the phase diagram in Fig. \ref{phase_diagram_potential} for which we provide here a
derivation. This also provides a derivation of the first part of the Letter by simply setting $a=0$. 
We compute the rank fields $r(x,t)$ and $s(x,t)$ at large times. Whenever there is a steady state, one can deduce the densities $\rho_s(x)$ and $\rho_d(x)$ using the parametric representation \eqref{rho_parametric} and its extension in all phases.

To test our predictions for large $N$ we have also performed numerical simulations by solving the Langevin equation 
and measuring the densities in the steady state for some finite values of $N$. The results are displayed in 
Fig \ref{fig_rhos_potential} and Fig \ref{fig_rhod_potential} and show excellent agreement with the predictions. 

\subsection{Attractive case} \label{app:potential_attractive}

We start with the attractive case. We first assume that there exists a steady state and solve the $T=0$ stationarity equations
\bea \label{system1_lin}
v_0 s' &=& 2 \bar \kappa r r' + a \, {\rm sgn}(x) \, r' \; , \\
v_0 r' &=& 2  \bar \kappa  r s' + a \, {\rm sgn}(x) \, s' - 2 \gamma s \;.\label{system2_lin}
\eea 
For $a>0$, the center of mass is attracted to the origin. It is thus natural 
to search for a stationary solution such that $r(x)$ is odd and $s(x)$ is even in $x$. 
For $a=0$ we use coordinates such that the center of mass is at zero, 
and thus it is also reasonable to assume such symmetry (note however that in the main text we did not make this assumption).
Thus we can restrict ourselves to $x\geq 0$ in the following, and impose the condition $r(0)=0$.
We proceed as in the text for $a=0$, by integrating the first equation with $r(+\infty)=1/2$ and $s(+\infty)=0$, to obtain for $x \geq 0$
\be \label{s_function_r_lin}
s(x) = \frac{\bar \kappa}{v_0} \big(r(x)^2 - \frac{1}{4} \big) + \frac{a}{v_0} \big(r(x) - \frac{1}{2} \big) \;.
\ee 
Note that by symmetry this equation holds for any $x \in \mathbb{R}$ if $r(x)$ in the last term is replaced by $|r(x)|$. 

Substituting in the second equation one obtains
\be \label{eqr_potential}
(v_0^2 - (2 \bar \kappa r + a)^2) r' 
= \frac{\gamma}{2} (1- 2 r) (2 a + \bar \kappa + 2 \bar \kappa r) \;.
\ee 
Let us now analyze this equation, starting with the case $\bar \kappa=0$.

{\bf Non-interacting case $\bar \kappa=0$.} Let us start with the non interacting case $\bar \kappa=0$. Equation \eqref{eqr_potential} becomes
\be \label{eqr_potentialNI}
(v_0^2 - a^2) r'
= a \gamma (1- 2 r) \;,
\ee 
a linear equation whose general solution for $x \geq 0$ is
\be 
1-2r(x) = C e^{- \frac{2 a \gamma x}{v_0^2-a^2}} \;.
\ee 
Since $r(+\infty) = \frac{1}{2}$ and $r(0)=0$, there are two phases:

(i) $a \geq v_0$, in which case $C=0$, leading to $r(x)=\frac{1}{2}$ for $x >0$
and, when extending to $x \in \mathbb{R}$
\be 
r(x)=\frac{1}{2} {\rm sgn}(x) \quad , \quad \rho_s(x) = \delta(x) 
\ee 
leading to $s(x)=0$.

(ii) $a<v_0$, in which case $C=1$ \cite{footnoteC}, leading to a smooth solution which reads, upon extending to $x \in \mathbb{R}$
\be \label{noninteracting0} 
r(x)=\frac{1}{2} \sgn(x) (1 - e^{- \frac{2 a \gamma x}{v_0^2-a^2}}) \quad , \quad \rho_s(x) = \frac{\gamma a}{v_0^2-a^2} \, \exp \left(-\frac{2\gamma a}{v_0^2-a^2}|x| \right)
\ee
where $s(x)$ is given for $x \geq 0$ by \eqref{s_function_r_lin}. 
We thus recover the known result of \cite{DKM19}.

{\bf Interacting case $\bar \kappa>0$}. Let us now focus on the attractive case $\bar \kappa >0$. The phase diagram in the space $(\bar \kappa/v_0,a/v_0)$ is similar to the case $a=0$ (the phases are the same, but the phase boundaries depend on $a/v_0$). 

{\bf Phase $I_s$}. For $v_0>a+\bar \kappa$, the prefactor on the l.h.s of 
\eqref{eqr_potential} is strictly positive. Since the r.h.s. is positive and bounded, we get that $r'(x)$ is bounded on $\mathbb{R}^+$. Hence the stationary density is smooth with infinite support, and $r(x)$ for $x \geq 0$ is obtained by inversion of the equation
\begin{equation}
\gamma x = {\sf f}_a(r)= 2 \bar \kappa r + \frac{v_0^2 - (\bar \kappa+a )^2 }{2(\bar \kappa+a)} \log\left( \frac{\bar \kappa + 2 a + 2 \bar \kappa r}{(\bar \kappa + 2 a)(1-2 r) } \right) \label{eqres_potential}
\end{equation}
and $s(x)$ is then given for $x \geq 0$ by \eqref{s_function_r_lin}. Note that for convenience we normalized so that ${\sf f}_{a=0}(r)=  \bar \kappa f(r)$, and
that ${\sf f}_a(0)=0$, ${\sf f}'_a(0)= 2 (v_0^2-a^2)/(2 a + \bar \kappa)$ and 
${\sf f}''_a(0)= 8 a (v_0^2-(a+\bar \kappa)^2)/(2 a + \bar \kappa)^2$.
This is the phase denoted $I_s$ in Fig. \ref{phase_diagram_potential}. The densities $\rho_s$ and $\rho_d$ have an infinite support.
They are smooth on their support except at $x=0$ for $a>0$. Let us now examine their behavior
near $x=0$. 

Let us assume, as is natural here, that $r(0^+)=0$ (no shock at zero). Then Eq. \eqref{eqr_potential} gives $r'(0^+)=\frac{\gamma (2 a + \bar \kappa)}{2 (v_0^2-a^2)}$.
Since $s'(0^+)=\frac{a}{v_0} r'(0^+)$ from \eqref{s_function_r_lin}, we obtain that for $x$ near zero
\be  \label{steprhod} 
\rho_d(x) \simeq \frac{a}{v_0} \frac{\gamma (2 a + \bar \kappa)}{2 (v_0^2-a^2)} {\rm sgn}(x) \;.
\ee 
Hence $\rho_d(x)$ has a step discontinuity at $x=0$. It means that there is an excess of $+$ particles for $x>0$ and of $-$ particles for $x<0$.
The result \eqref{steprhod}  is valid with the phases $I_s$ and $F_e$. It is also true for the non-interacting case $\bar \kappa=0$. 
Note that the discontinuity diverges as $v_0 \to a^-$. 

Now, it is easy to see that for $a>0$, $\rho_s(x)$ has also a non-analyticity at $x=0$. Indeed inverting \eqref{eqres_potential} near $x=0$ 
one finds (taking into account that $r(x)$ must be odd) 
\be 
r(x) = \frac{\gamma x}{{\sf f}'_a(0)} - \frac{{\sf f}''_a(0)}{2 {\sf f}'_a(0)^3} \gamma^2 x^2 {\rm sgn}(x) + O(x^3) 
\ee 
which leads to
\be \label{r_IsFe_cusp}
r(x) = \frac{2 a + \bar \kappa}{2 (v_0^2 - a^2) } \gamma x - \frac{a (2 a + \bar \kappa) (v_0^2 - (a + \bar \kappa)^2) }{2 (v_0^2 -a^2)^3 } \gamma^2 x^2 {\rm sgn}(x) + O(x^3)
\ee 
and finally one obtains for $x$ near $0$
\be \label{rhos_IsFe_cusp}
\rho_s(x) = \frac{2 a + \bar \kappa}{2 (v_0^2 - a^2) } \gamma - \frac{a (2 a + \bar \kappa) (v_0^2 - (a + \bar \kappa)^2) }{ (v_0^2 -a^2)^3 } \gamma^2 |x| 
+ O(x^2)
\ee 
which shows that $\rho_s(x)$ is continuous but with a derivative discontinuity at $x=0$. 
In this phase the density is maximum at the origin.

{\it In the critical case $v_0 = \bar \kappa + a$}, i.e. at the phase boundary $I_s$-$F_e$, one has simply 
${\sf f}_a(r)=2 \bar \kappa r$ and \eqref{eqres_potential} leads to exactly
the same linear solution \eqref{v0=kappa_solr} for $r(x)$ than for $a=0$, hence one obtains
\be \label{v0=kappa_solr2}
r(x) = \frac{1}{2} \frac{x}{x_e^*} \quad , \quad  s(x) = \frac{\bar \kappa}{4 v_0} \big((\frac{x}{x^*_e})^2 - 1\big) + \frac{a}{2 v_0} (\frac{|x|}{x_e^*}-1) 
\quad , \quad |x| \leq x_e^*= \frac{\bar \kappa}{\gamma} 
\ee 
which lead to the densities
\be \label{v0=kappa_solrho2}
\rho_s(x) = \frac{\gamma}{2\bar \kappa} \quad , \quad \rho_d(x) = \frac{ \gamma^2 x}{2 v_0 \bar \kappa} + \frac{a \gamma}{2 v_0 \bar \kappa} {\rm sgn}(x) 
\quad , \quad |x| \leq x_e^*= \frac{\bar \kappa}{\gamma} 
\ee 
which vanish for $|x|>x_e^*$. They have step discontinuities at the two edges, and in addition $\rho_d(x)$
has a step discontinuity at the origin, as discussed above.

{\bf Phase $F_e$}. For $v_0<\bar \kappa + a$, ${\sf f}_a(r)$ becomes non-invertible (recall that ${\sf f}_0(r)$ is plotted in Fig. \ref{fig_attractive}). The density then has a finite support with shocks at the edges $\pm x_e$. As in the case $a=0$, integrating equations (\ref{system1_lin}-\ref{system2_lin}) around $x_e>0$ with the correct interpretation of the force term as discussed in the text leads to
\be 
v_0 \Delta s = \left( \bar \kappa (r(x_e^+)+r(x_e^-)) + a \right)  \Delta r  ~,~ 
v_0 \Delta r = \left( \bar \kappa (r(x_e^+)+r(x_e^-)) + a) \right) \Delta s  
\ee 
with $\Delta r=r(x_e^+)-r(x_e^-)$ and $\Delta s=s(x_e^+)-s(x_e^-)$. A non-zero $\Delta r$ then implies $r(x_e^+)+r(x_e^-)=(v_0-a)/\bar \kappa$, leading to
\be \label{rxe} 
r(x_e^-) = \frac{v_0-a}{\bar \kappa} - \frac{1}{2}
\ee
since $r(x_e^+)=1/2$. Inside the support, $r(x)$ is again given by \eqref{eqres_potential}, where the
function ${\sf f}_a(r)$ is invertible on the interval $[0,r(x_e^-)]$. Indeed the point where
${\sf f}'_a(r)=0$ is $r=r^*= (v_0-a)/(2 \kappa) > r(x_e^-)$. 
This is the phase denoted $F_e$ in Fig. \ref{phase_diagram_potential}. 
The densities $\rho_s$ and $\rho_d$ have a bounded support $[-x_e,x_e]$
with delta singularities at the two edges, i.e one has for $x \in \mathbb{R}$
\be
\rho_s(x) = \Delta r (\delta(x-x_e) + \delta(x+x_e)) + \tilde \rho_s(x) ~ , ~ \rho_d(x) = \Delta s (\delta(x-x_e) - \delta(x+x_e)) + \tilde \rho_d(x) 
~ , ~ \Delta r = \Delta s = 1-\frac{v_0-a}{\bar \kappa} 
\ee
where $\tilde \rho_s$ and $\tilde \rho_d$ are smooth functions 
on $]-x_e,0[$ and $]0,x_e[$ (since ${\sf f}_a(r)$ is generically
analytic near $r=r(x_e^-)$ it implies that $\tilde \rho_s(x_e^-)$ and $\tilde \rho_d(x_e^-)$ are finite, together with all their left derivatives at $x=x_e^-$).
This corresponds to a cluster of $+$ particles at $x=x_e$
and a cluster of $-$ particles at $x=-x_e$. This is the phase denoted $F_e$ in Fig. \ref{phase_diagram_potential}.
Since inside the support the formulae for $r(x),s(x),\rho_s(x)$ and $\rho_d(x)$ are the same as for the phase $I_s$ above,
one finds that their behavior around $x=0$ is again given by \eqref{steprhod} and \eqref{rhos_IsFe_cusp}.
Hence in this phase $\rho_d(x)$ has also a jump at $x=0$ and $\rho_s(x)$ has a linear cusp,
with however positive amplitude, i.e. the density $\rho_s(x)$ is minimum at $x=0$.

{\bf Phase $\delta_0$}. 
For $v_0 \leq \bar \kappa/2 + a$ all particles are in the same cluster. 
Indeed from \eqref{rxe} we see that as $v_0 \to \bar \kappa/2 + a$, $r(x_e^-) \to 0$
implying $x_e \to 0$. This is the phase denoted $\delta_0$ in Fig. \ref{phase_diagram_potential}.
In that phase $r(x)=\frac{1}{2} {\rm sgn}(x)$ and $s(x)=0$, leading to 
$\rho_s(x)=\delta(x)$ and $\rho_d(x)=0$. 

{\bf Remark: stability of a single cluster at finite $N$}. Let us recall that for $a=0$ and for $v_0 < \bar \kappa/2$ there is a stability argument (given
in the text) for all particles to be in a single cluster for arbitrary $N$. Let us now extend this argument for $a>0$. 
Denote $x_+$ (resp. $x_-$) the position of one of the rightmost (resp. leftmost) particles
and $n_+$ (resp. $n_-$) the number of particles at the same location.
One has 
\bea \label{eqx+}
&& \frac{dx_+}{dt} \leq v_0 - \frac{\bar \kappa}{N} (N-n_+) - a \sgn(x_+)
\\
&& \frac{dx_-}{dt} \geq -v_0 + \frac{\bar \kappa}{N} (N-n_-) - a \sgn(x_-)
\eea
which leads to
\be
\frac{d(x_+-x_-)}{dt} \leq 2 v_0 - \bar \kappa \left( 2-\frac{n_++n_-}{N} \right) - a (\sgn(x_+) -\sgn(x_-)) 
\leq 2v_0 - \bar \kappa
\ee
since one has $\sgn(x_+) -\sgn(x_-) \geq 0$. This implies that for $v_0 < \bar \kappa/2$ 
all particles will end up in a single cluster. Note however that for finite $N$ the $\delta$ peak in the density is not necessarily at $x=0$ in this case. On the other hand if $v_0<a$, then as long as $x_+>0$, $\frac{dx_+}{dt} < v_0-a < 0$, 
and symmetrically for $x_-$. This also implies, for any $N$, that all particles will end up in a single cluster,
which in this case must be at $x=0$. Thus, a delta peak in the density $\rho_s(x)$ containing all the particles 
is stable for any $N$ for $v_0<\min(a,\bar \kappa/2)$. However in the regime $\min(a,\bar \kappa/2)<v_0<\bar \kappa/2+a$, it is stable only in the limit $N \to + \infty$, while for finite $N$ there can be deviations from this state. Consider for example that all $N$ particles are at $x=0$, all in the $+$ state. Then they will have a total velocity $v_0-a>0$ and will be able to escape from $x=0$. Imagine now that at some point half of the particles switch sign. Then the cluster will break since $v_0>\bar \kappa/2$. As $N\to+\infty$, the probability of such events decreases exponentially, leading to the $\delta_0$ phase.


\begin{figure}
    \centering
    \includegraphics[width=0.32\linewidth]{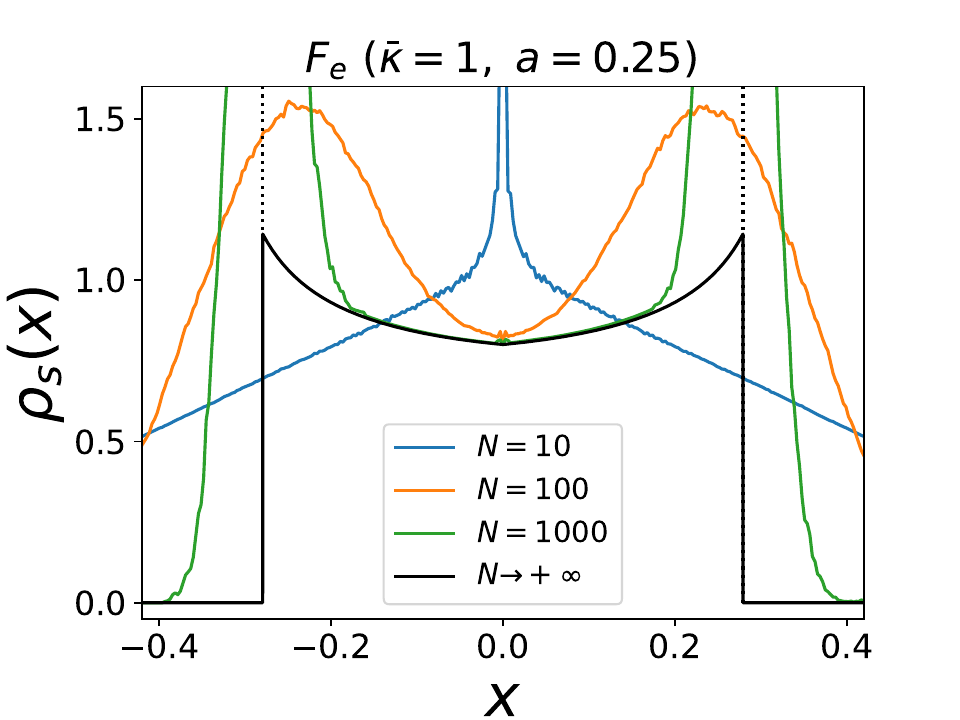}
    \includegraphics[width=0.32\linewidth]{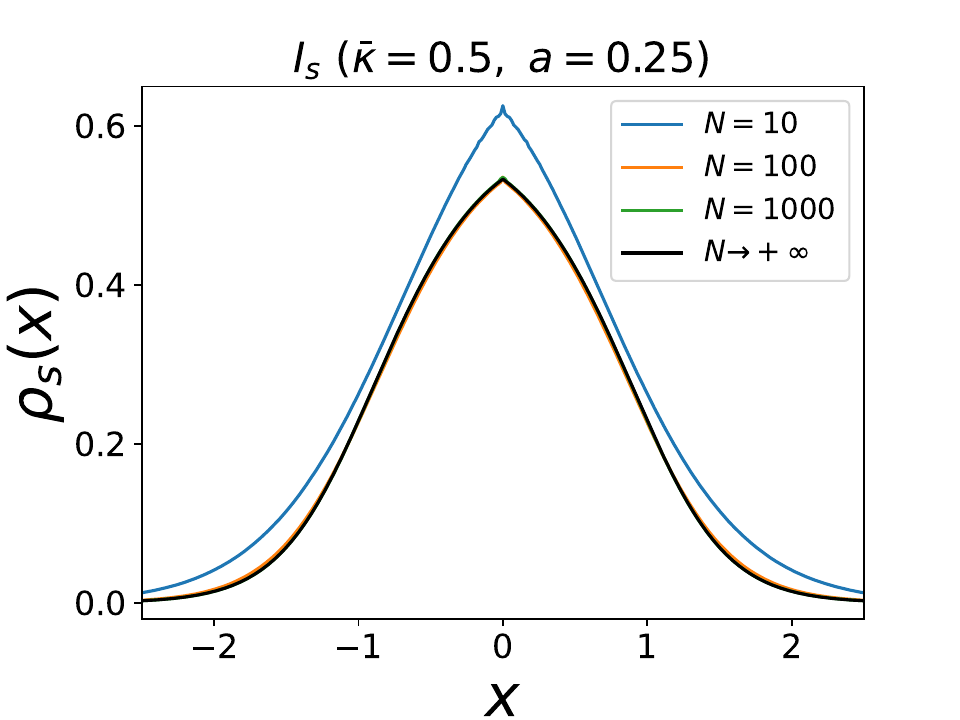}
    \includegraphics[width=0.32\linewidth]{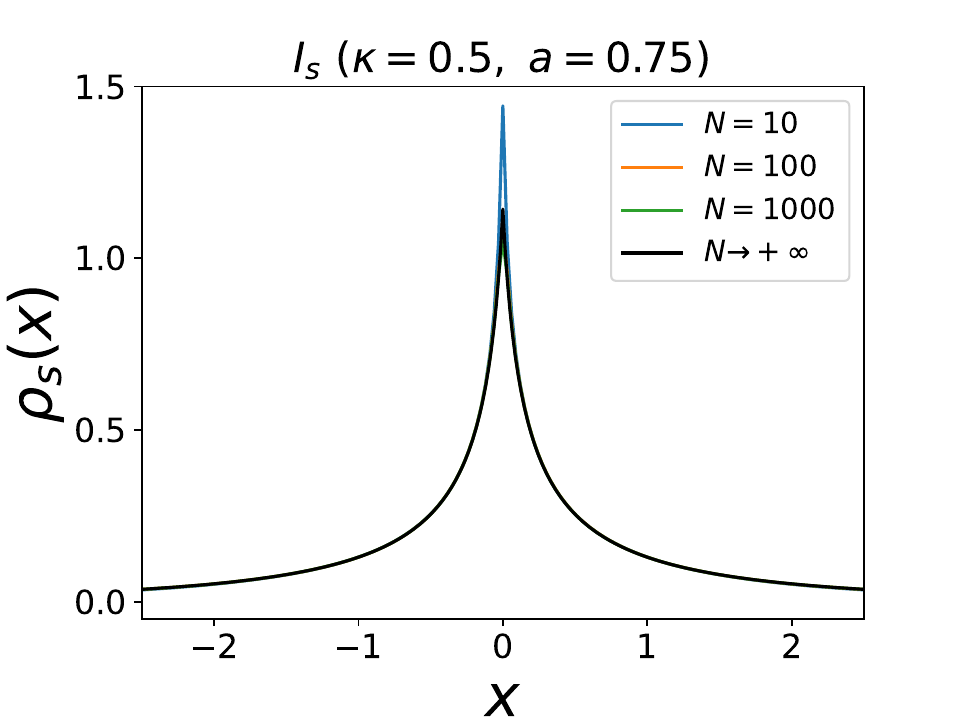}
    \includegraphics[width=0.32\linewidth]{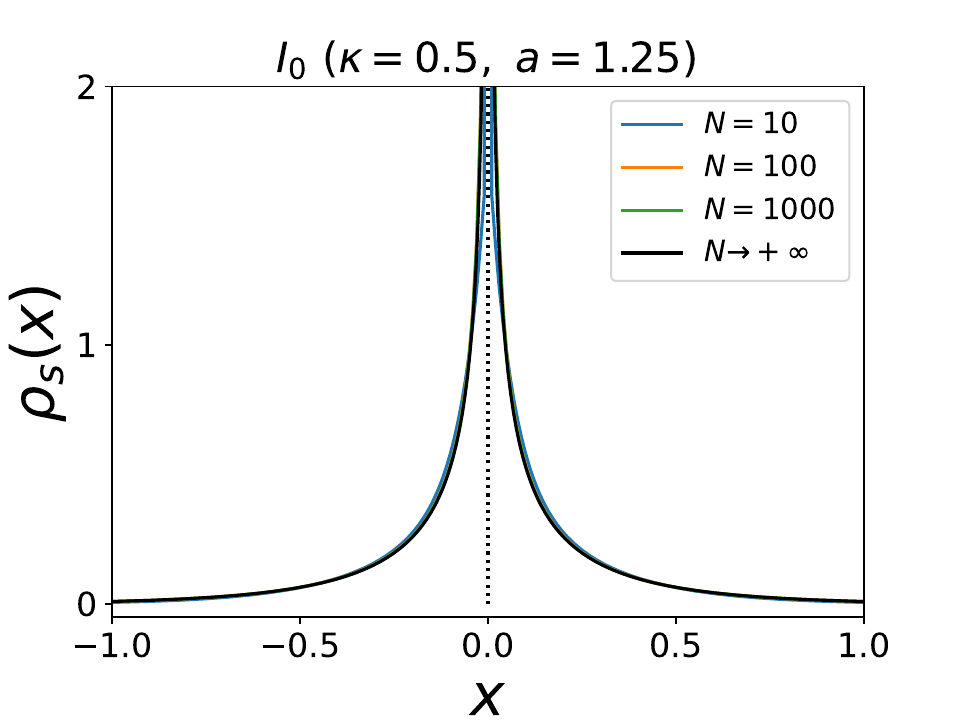}
    \includegraphics[width=0.32\linewidth]{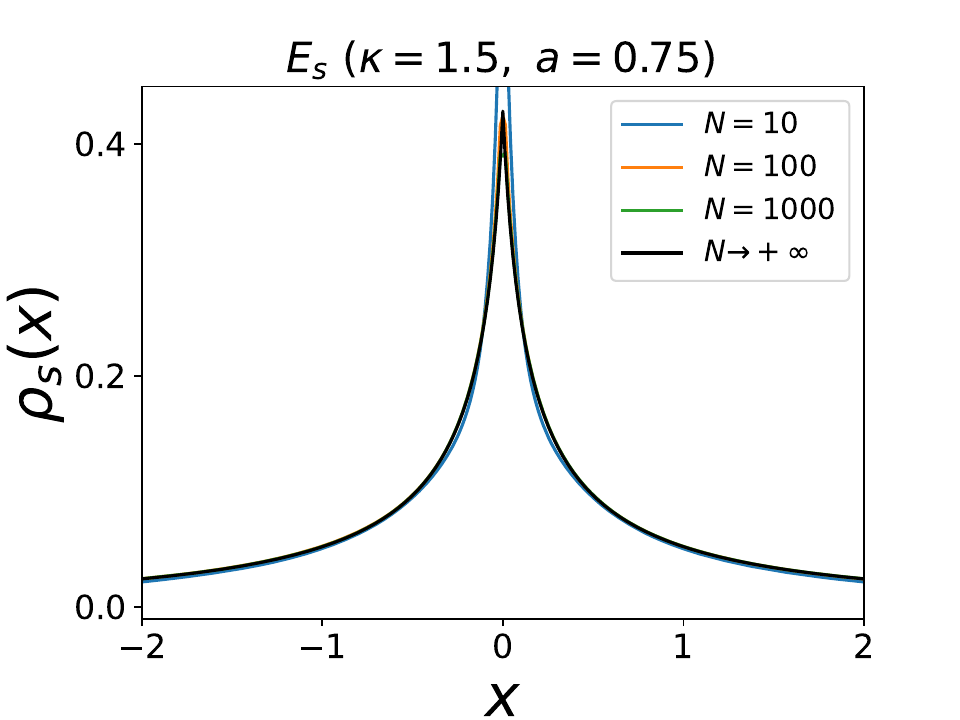}
    \includegraphics[width=0.32\linewidth]{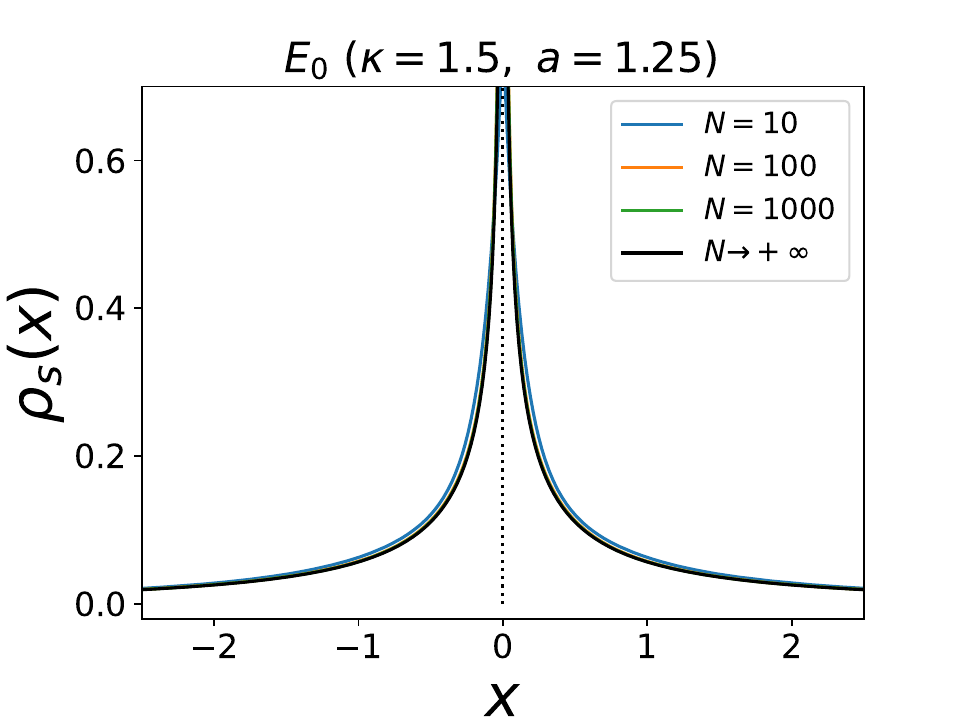}
    \caption{Density $\rho_s(x)$ in the steady state in the different phases, in the presence of a potential $V(x)=a|x|$, for different values of $N$. For all figures $v_0=1$ and $\gamma=1$. For finite $N$ it is obtained by numerical simulations (note that here we have not made the shift to the reference frame of the center of mass contrary to the case $a=0$ discussed in the main text). The black curves correspond to the limit $N\to +\infty$ and are computed using the parametric representation \eqref{rho_parametric} extended to all phases. The delta peaks in the density are represented by dotted vertical lines. Note that in all phases except $F_e$ the data for $N=100$ is already hardly distinguishable from the prediction at $N=+\infty$.}
    \label{fig_rhos_potential}
\end{figure}

\subsection{Repulsive case}
We now turn to the repulsive case, $\kappa =-\bar \kappa >0$. While for $a=0$, there is a single, expanding phase, the addition of a linear potential makes the phenomenology much richer: for $a>0$, 5 different phases can be distinguished. Two of these phases ($I_s$ and $\delta_0$) are continuations
of phases which exist for $\bar \kappa>0$.

{\bf Phase $\delta_0$}. 
First of all, when $a\geq v_0+\kappa$, the density trivially converges to $\rho_s(x)=\delta(x)$ independently of $N$ (as can be seen e.g. from
\eqref{eqx+} setting $\bar \kappa=-\kappa$). This is again the phase $\delta_0$ in Fig. \ref{phase_diagram_potential}, but
on the side $\bar \kappa<0$.

For $a<v_0+\kappa$, one needs to distinguish between a stationary case for $\kappa\leq a<v_0+\kappa$, and an expanding case for $a<\kappa$.
Each of these cases will give rise to two distinct phases, separated by the line $a=v_0$, leading to four phases in total. 
Let us start with the stationary case, where we again solve Eqs. (\ref{system1_lin}-\ref{system2_lin}), but this time for $\bar \kappa=-\kappa<0$.
We can again look for solutions with $r(x)$ odd in $x$ and $s(x)$ even.

{\bf Phase $I_s$}. 
For $\kappa = -\bar \kappa >0$, \eqref{eqr_potential} becomes (for $x \geq 0$)
\be \label{eqkappa} 
(v_0^2 - (a - 2\kappa r)^2) r' = \frac{\gamma}{2} (1- 2r) (2 a - \kappa - 2 \kappa r) \;.
\ee 
When $a > \kappa$, the right hand side is always positive. If in addition $a < v_0$, then the factor in the left-hand side is strictly positive and $r(x)$ is correctly given by equation \eqref{eqres_potential} for any $x\geq 0$. The function ${\sf f}_a(r)$ is invertible, 
and this is the same phase $I_s$ as discussed above: both densities $\rho_s(x)$ and $\rho_d(x)$ are smooth
with infinite support.

{\bf Phase $I_0$}. 
However, if $a>v_0$ (still assuming $\kappa\leq a<v_0+\kappa$) the left-hand side of \eqref{eqkappa} is positive only above a certain value of $r$. Thus, for 
$r'(x)$ to be positive everywhere there must be a jump at $x=0$, such that
\be \label{jump0}
r(0^+) = \frac{a-v_0}{2\kappa} \;.
\ee
and $r(x)$ for $x>0$ is now given by
\be
\gamma x = {\sf f}_a(r) - {\sf f}_a(r(0^+))
\ee
where ${\sf f}_a(r)$ is still given by \eqref{eqres_potential}. This the phase $I_0$ in Fig. \ref{phase_diagram_potential},
which is stationary, has infinite support and exhibits a shock at $x=0$. Note also that $\rho_s(x)=r'(x)$ diverges near $x=0$. More
precisely near $x=0$ one has, by linearizing \eqref{eqkappa} for $r$ near $r(0^+)$ and integrating w.r.t. $x > 0$,
\be 
2 \kappa v_0  (r(x)-r(0^+))^2 \sim 
\frac{\gamma}{2 \kappa} (v_0^2 - (a- \kappa)^2) \, x \;.
\ee 
Hence there is generically an inverse square root divergence of the density $\rho_s(x)$ near the shock at $x=0$, i.e. for $x \in \mathbb{R}$
and $x$ near $0$ one has
\be 
r(x) \simeq \left( r(0^+)  + B \sqrt{|x|} \right)  {\rm sgn}(x)  \quad , \quad \rho_s(x) \simeq \frac{a-v_0}{\kappa}  \delta(x) + \frac{B}{2 \sqrt{|x|}} \quad {\rm with} \quad B = \frac{1}{2\kappa} \sqrt{\frac{\gamma}{v_0} (v_0^2 - (a- \kappa)^2)} \;. \label{sqrtdiv} 
\ee 
Using \eqref{s_function_r_lin} and the value of $r(0^+)$ in \eqref{jump0}, we find that
\be 
s(x) \simeq \frac{(a-\kappa)^2-v_0^2}{4 \kappa v_0} + B \sqrt{|x|} \label{sxI0}
\ee 
and since $s(x)$ is even, $s(x)$ is continuous at $x=0$ which implies that the cluster of particles there
has equal fractions of $\pm$ species. However, Eq. \eqref{sxI0} leads to an inverse square root divergence of the density $\rho_d(x)$ 
\be 
\rho_d(x) \simeq  \frac{B}{2 \sqrt{|x|}} \, {\rm sgn}(x) \;.
\ee 
This result shows that the inverse square root divergence involves only $+$ particles for $x>0$ and
minus particles for $x<0$. 

The presence of the cluster of
particles at $x=0$ can easily be understood via a qualitative argument. When $v_0<a$, non-interacting particles would remain stuck at $x=0$, and it is only the repulsion that allows some particles to escape from $x=0$. Therefore there must be a cluster of particles at $x=0$, which contains a fraction $r(0^+)-r(0^-)=2r(0^+)$ of particles. The particle immediately at the right of this cluster is able to escape if the total force resulting from the potential and the interactions $-a+2\kappa r(0^+)$ is larger than $-v_0$. Balancing these two forces fixes the value of $r(0^+)$ in agreement with \eqref{jump0}. The presence of the square root divergence in the density near the cluster comes from the fact that the velocity of particles decreases to zero as one approaches the cluster.

Note that as one gets closer to the phase boundary $I_0 - \delta_0$, i.e. $a - \kappa \to v_0$, the weight of the
$\delta$ peak in $\rho_s(x)$ tends to unity and $B \to 0$. 

{\it Large distance behavior in $I_s$ and $I_0$ phases}. 

Let us first note that the behaviors at large $|x|$ of $\rho_s(x)$ and $\rho_d(x)$ are related in a simple way.
Indeed, 
Eq. \eqref{s_function_r_lin} implies that whenever the support of the densities in infinite one has as $|x| \to + \infty$
\be 
\rho_d(x) \simeq \frac{a + \bar \kappa}{v_0} \rho_s(x) {\rm sgn}(x) \;.
\ee 
One can check that in the phases $I_s$ and $I_0$ the prefactor is smaller than unity, as it should.
One sees that as $v_0$ increases the proportion of $-$ on the positive side increases and vice-versa, i.e. the mixing of $+$ and $-$ particles becomes stronger. 

Next, let us determine this large distance behavior. Expanding Eq. \eqref{eqres_potential} for $r$ near $1/2$ and $x \to +\infty$ one finds
\be 
r(x) = \frac{1}{2} - \frac{a+\bar \kappa}{2 a + \bar \kappa} e^{- \frac{2  (a+\bar \kappa)}{v_0^2 - (a+\bar \kappa)^2} (\gamma x - \bar \kappa)  } 
\ee 
which recovers \eqref{noninteracting0} for $\bar \kappa=0$. This leads to
\bea \label{asymptotics2} 
&& \rho_s(x) \simeq A_s \, e^{- \frac{|x|-x_e^*}{\xi_\infty}}  ~,~ \rho_d(x) \simeq A_d \, {\rm sgn}(x) e^{- \frac{|x|-x_e^*}{\xi_\infty}} \\
&& \xi_\infty =  \frac{v_0^2 - (\bar \kappa+a)^2}{ 2 \gamma (\bar \kappa+a)  } ~,~ x_e^*=\frac{\bar \kappa}{\gamma}  ~,~  A_s = \frac{ 2 \gamma (\bar \kappa+a)^2}{(v_0^2 - (\bar \kappa+a)^2) (\bar\kappa+2a)} ~ , ~ A_d=  \frac{a+\bar \kappa}{v_0} A_s \nonumber \;. 
\eea 
which for $a=0$ reproduces the result given in the text. This is valid within the phases $I_s$ and $I_0$ both
in attractive and repulsive case with $\bar \kappa = - \kappa$. 

{\it Critical case $a=\kappa$}. This corresponds to the boundaries $I_s-E_s$ and $I_0-E_0$. The reasoning above still holds in the marginal case $a=\kappa$, both for 
$a\leq v_0$ (boundary $I_s-E_s$) and for $a>v_0$ (boundary $I_0-E_0$).
Eq. \eqref{eqres_potential} takes a much simpler form in this case, 
\bea \label{eqres_potential_akappa}
&& \gamma x = \tilde {\sf f}_a(r)= 2 a r \left( \frac{v_0^2}{a^2(1-2r)} - 1 \right)  \quad , \quad a< v_0 \quad , \quad x \geq 0 \\
&& \gamma x = \tilde {\sf f}_a(r) - \tilde {\sf f}_a(r(0^+)) = \frac{(v_0-a + 2 a r)^2}{a(1-2 r)}
\quad , \quad a > v_0 \quad , \quad x > 0 \;.
\eea 

This leads to
\bea 
&& r(x) = \frac{1}{4 a^2} \left( \sqrt{ (a \gamma x + (a-v_0)^2) (a \gamma x + (a+v_0)^2) } +a^2  - v_0^2 - a \gamma x \right)  \quad , \quad a< v_0 
 \quad , \quad x \geq 0 \\
&& r(x) = \frac{1}{4 a} \left(   \sqrt{ \gamma x (4 v_0 + \gamma x) } + 2 (a - v_0) - \gamma x \right) 
\quad , \quad a > v_0 \quad , \quad x > 0
\eea  

and, for $x \in \mathbb{R}$ 
\bea  \label{rhosEs}
&& \rho_s(x) = \frac{\gamma}{4 a} \left( \frac{a^2 + v_0^2 + a \gamma |x|}{\sqrt{ (a \gamma |x| + (a-v_0)^2) (a \gamma |x| + (a+v_0)^2) }} -1 \right) 
\quad , \quad a< v_0  \\
&&  \rho_s(x) = \frac{\gamma}{4 a} \left(  \frac{2 v_0 + \gamma |x|}{\sqrt{ \gamma |x| (4 v_0 + \gamma |x|) }} - 1 \right) + \left(1-\frac{v_0}{a}\right) \delta(x)
\quad , \quad a > v_0 \;. \label{rhosE0}
\eea   
We note that for $a>v_0$ the behavior of the density near the shock at $x=0$ is compatible with
the more general result \eqref{sqrtdiv} with an inverse square root divergence of amplitude $B= \sqrt{\gamma v_0}/(2 a)$. 
Finally, an unusual feature is the algebraic decay of the density at infinity. This can be
anticipated since the bound state decay length $\xi_\infty$ in \eqref{asymptotics2}  diverges as $\kappa \to a^-$.
Indeed, one finds at large $x$ in both cases
($a<v_0$ and $a>v_0$) 
\be 
\rho_s(x) \simeq \frac{v_0^2}{2 a \gamma x^2} 
\ee 
where only the term $O(1/|x|^3)$ differs for $a<v_0$ and $a>v_0$. 
Note that for $a=\kappa$ one has
\be 
s(x) = - \frac{a}{4 v_0} (1- 2 r(x))^2 \;.
\ee 
This implies that at large $x>0$ one has $s(x) \simeq - \frac{1}{4} \frac{v_0^3}{a \gamma^2 x^2}$, leading to
\be 
\rho_d(x) \simeq \frac{v_0^3}{2 a \gamma^2 x^3} \;.
\ee

\begin{figure}
    \centering
    \includegraphics[width=0.32\linewidth]{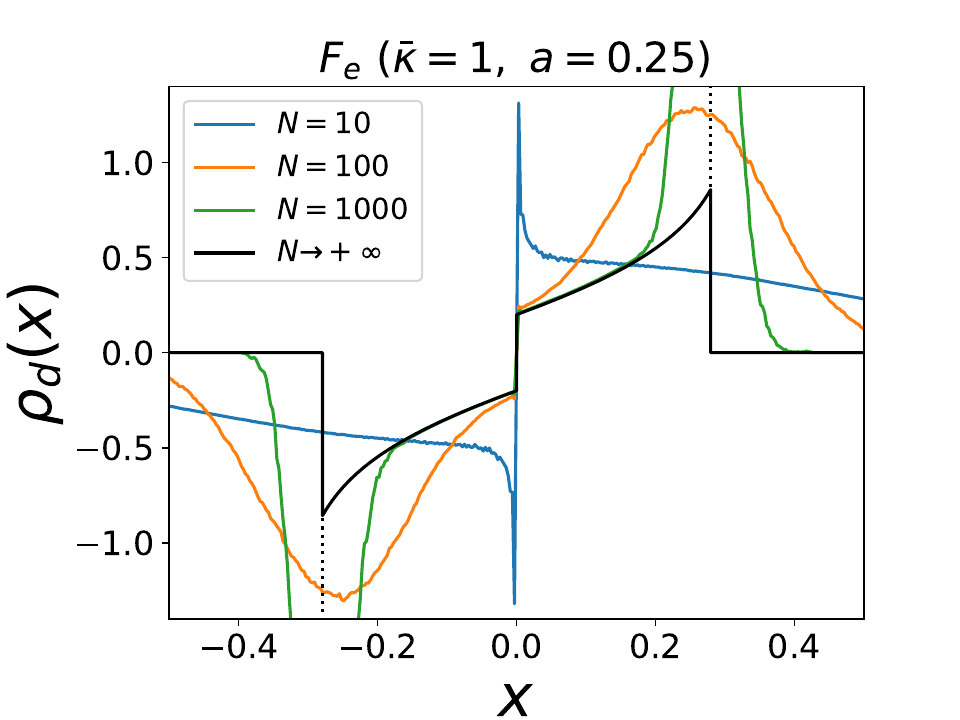}
    \includegraphics[width=0.32\linewidth]{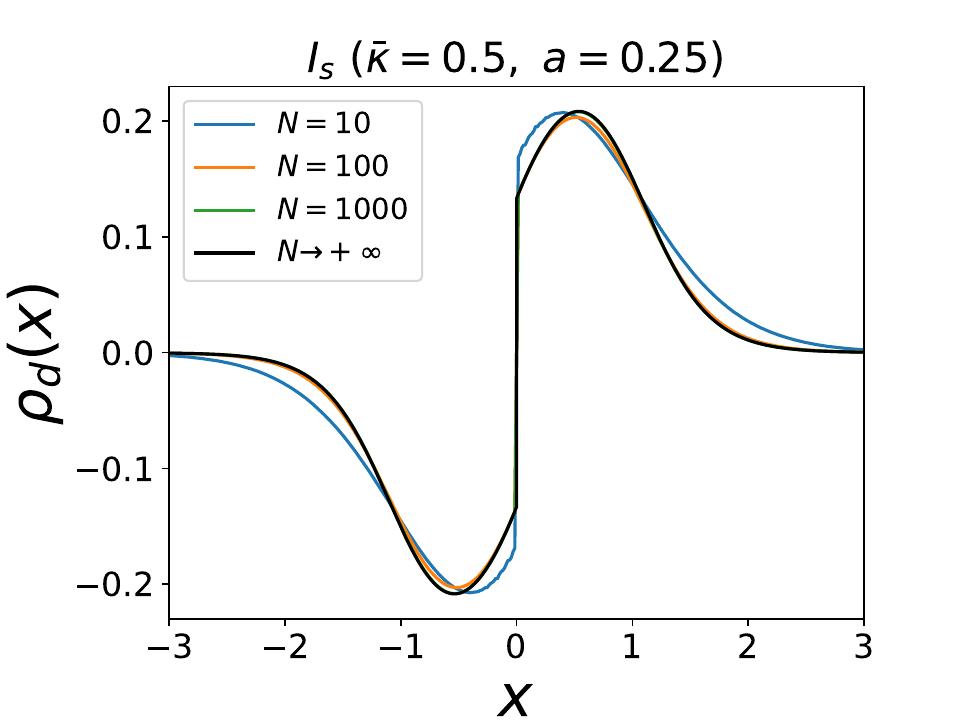}
    \includegraphics[width=0.32\linewidth]{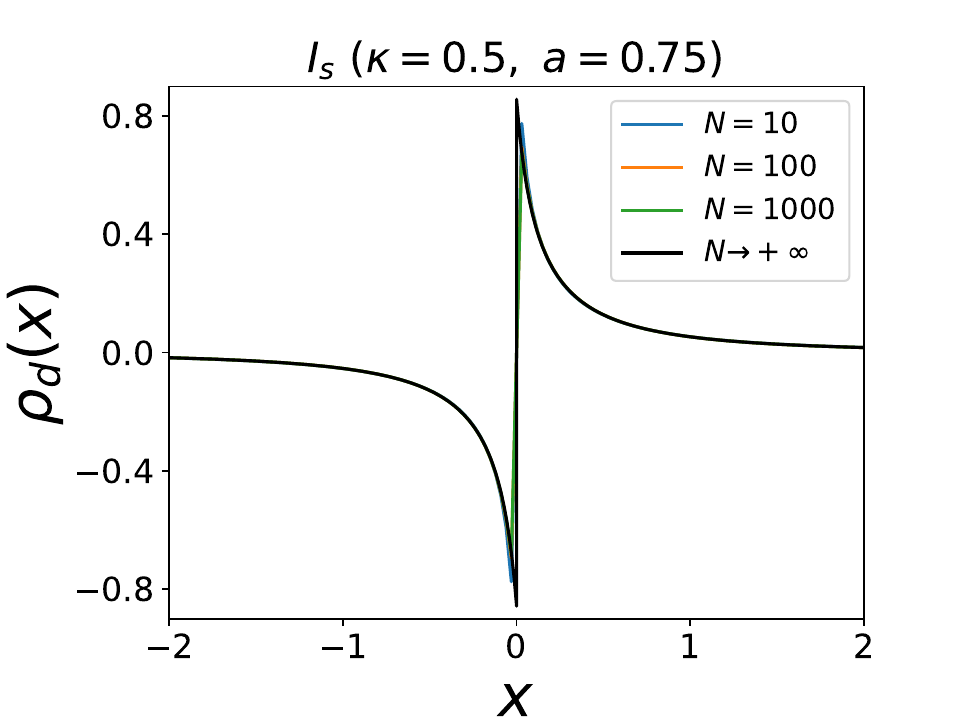}
    \includegraphics[width=0.32\linewidth]{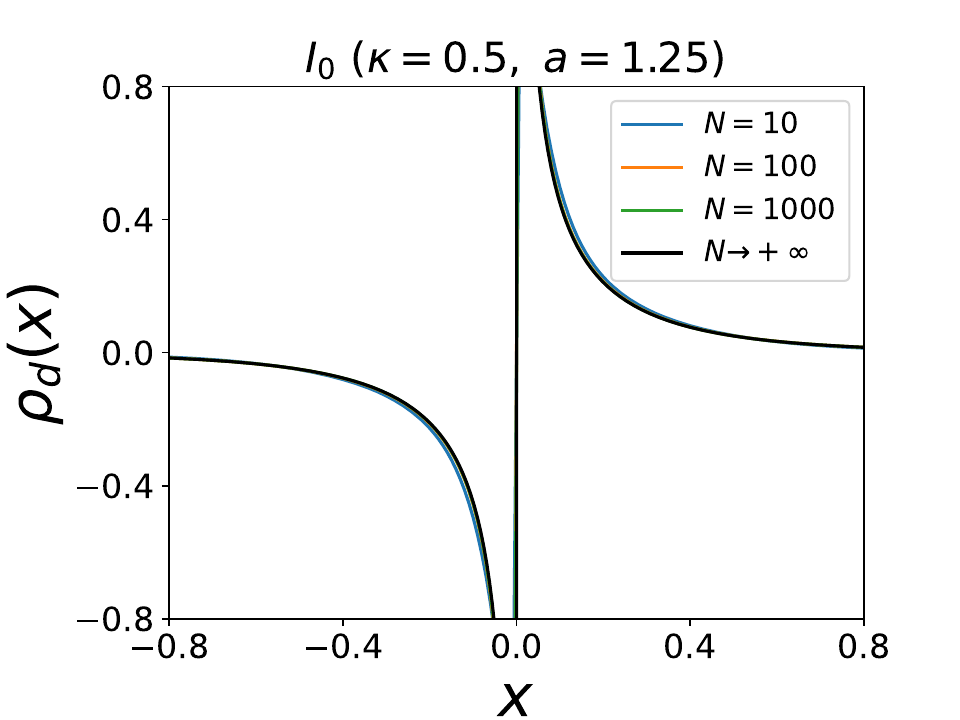}
    \includegraphics[width=0.32\linewidth]{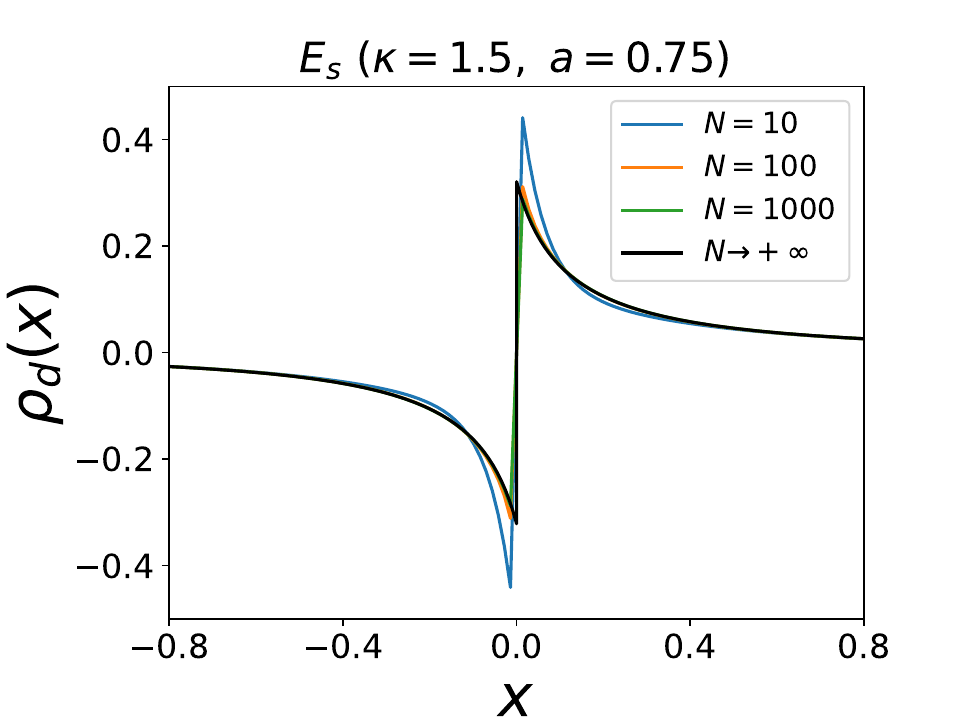}
    \includegraphics[width=0.32\linewidth]{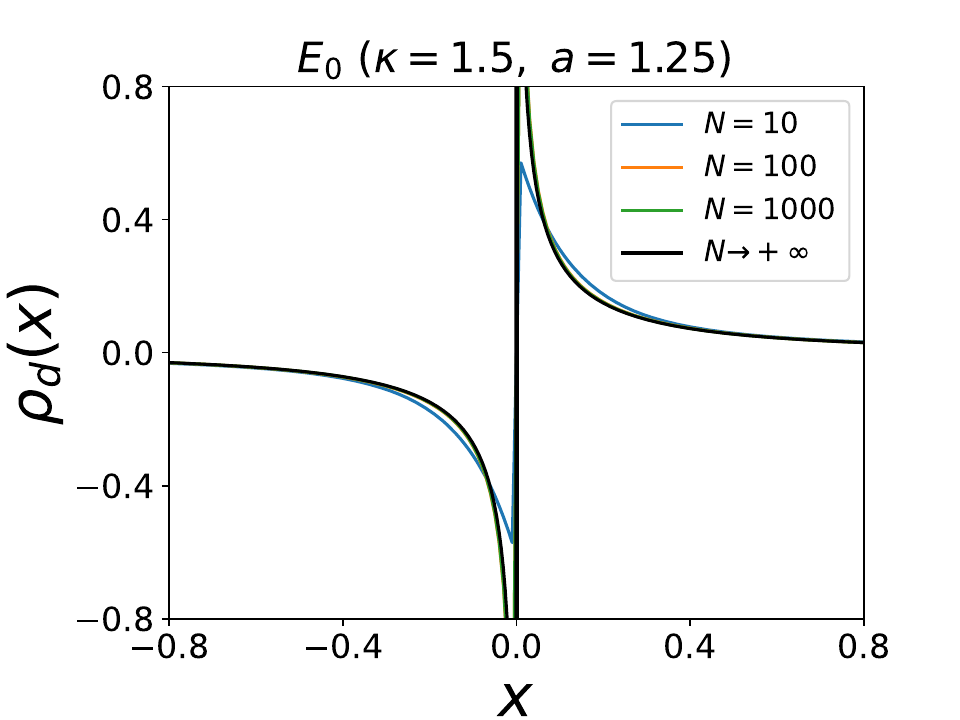}
    \caption{Density $\rho_d(x)$ in the steady state in the different phases, in the presence of a potential $V(x)=a|x|$, for different values of $N$. 
    For finite $N$ it is obtained by numerical simulations (note that here we have not made the shift to the reference frame of the center of mass contrary to the case $a=0$ discussed in the main text).
    For all figures $v_0=1$ and $\gamma=1$. The black curves correspond to the limit $N\to +\infty$ and are computed using the parametric representation \eqref{rho_parametric} and its extension to all phases. The delta peaks in the density are represented by dotted vertical lines.}
    \label{fig_rhod_potential}
\end{figure}

{\bf Expanding phases $E_s$ and $E_0$.}
Let us now turn to the expanding case $\kappa>a$, which contains two phases separated by the line $v_0=a$. In this case the repulsion is strong enough to overcome the potential and a fraction of the particles is sent to infinity. Thus at large time, there are two regions of interest in space.

(i) First, the particles which escape to infinity 
will form an expanding gas on scales $|x| \sim t$, with two edges moving at a velocity $\pm (\kappa-a)$,
each with an associated boundary layer of spatial width $\sim \sqrt{t}$. Below we determine the large time densities which 
we find identical for the two phases $E_s$ and $E_0$. 

(ii) Next the particles which remain inside the potential well will exhibit a stationary density on distances of order $O(1)$,
which we will compute. This density will either exhibit a shock at $x=0$ (phase $E_0$) or will be smooth (phase $E_s$),
in both cases with infinite support.
\\

{\it (i) Escaping particles.} Let us start with the escaping particles. 
As for $a=0$, we start from the equations
\bea \label{system1_potential_time}
\partial_t r &=& T \partial_x^2 r -v_0 \partial_x s - 2 \kappa r \partial_x r + a \, {\rm sgn}(x) \partial_x r \;, \\
\partial_t s &=& T \partial_x^2 s -v_0 \partial_x r - 2 \kappa r \partial_x s + a \, {\rm sgn}(x) \partial_x s - 2 \gamma s \;,
\label{system2_potential_time}
\eea
where we have reintroduced the temperature since it does not make the derivation more difficult in this case. We define $y=x/t$ and search for a solution of the form
\be 
r(x,t)= r_0(y)  + \frac{r_1(y)}{t} + \dots \quad , \quad 
s(x,t) = \frac{s_1(y)}{t}  + \dots
\ee 
The first equation \eqref{system1_potential_time} gives, at leading order in $1/t$
\be
(2 \kappa r_0 - y - a \, \sgn(y)) r_0' = 0 \;.
\ee
Thus for a given $y$ either $r_0'(y)=0$ or $r_0(y)=(y+a \, \sgn(y))/(2\kappa)$. Since $|r_0(y)|\leq 1/2$ we obtain the leading behavior of the 
solution for $x>0$ in the scaling region $x \sim t$ 
\be
r(x,t) \simeq \min \left(\frac{a}{2\kappa} + \frac{x}{2\kappa t}, \frac{1}{2} \right) \;.
\ee
By symmetry this gives the following density for $x \in \mathbb{R}$ at large times in the scaling region $|x| \sim t$
\be
\rho_s(x,t) \simeq \frac{a}{\kappa} \delta(x) + \frac{1}{2\kappa t} \theta((\kappa-a)t - |x|) \;, 
\ee
i.e. uniform on the support $[-(\kappa-a)t,(\kappa-a)t]$ with a "cluster" at $\frac{x}{t}=y=0$ containing a fraction
\be 
2 r_0(y=0^+)= \frac{a}{\kappa}
\ee
of particles.
Our solution only shows that these particles have zero velocity on average, 
and have remained in the vicinity of the potential well. They do not
however necessarily form a real cluster
in the sense encountered above. Below we will study the structure of their density 
on scales $|x| \ll t$. 

The second equation \eqref{system2_potential_time} gives, at leading order in $1/t$ the same relation as for $a=0$, namely
\be \label{r_s_potential_expanding}
s_1=-\frac{v_0}{2\gamma}r_0' \;,
\ee
hence, using the solution for $r_0(y)$ one obtains, in the scaling region $x \sim t$
\be
s(x,t) \simeq -\frac{v_0}{4\gamma\kappa t} \theta((\kappa-a)t - |x|)
\ee
and thus
\be
\rho_d(x,t) \simeq \frac{v_0}{4\gamma\kappa t} (\delta((\kappa-a)t - x) - \delta((\kappa-a)t - x)) \;. 
\ee
As for $a=0$, \eqref{r_s_potential_expanding} shows that at large times, the noise plays the same role as a Brownian noise with temperature $T_a=\frac{v_0^2}{2\gamma}$, and one can for instance compute the boundary layer in exactly the same way (see below).

The present expansion is formally correct inside the "bulk" of the expanding gas, i.e. for $|y|=\frac{|x|}{t} < \kappa-a$.
However examining the equations (\ref{system1_potential_time}-\ref{system2_potential_time}) to the next orders in $1/t$,
one finds an indeterminacy, e.g. $r_1(y)$
cannot be determined. The reason for that is that the higher order terms, starting with $r_1(y)$
may depend on the details of the initial condition. One can see how this happens in the simpler
case of the Burger's equation with $a=0$, which amounts to set $v_0=0$ in \eqref{system1_potential_time} and $T=0$. In
that case the solution $r(x,t)$ starting from the initial condition $r(x,0)=R(x)$ is given
by the solution of the equation 
\be 
r = R(x- 2 \kappa r t) \quad \Leftrightarrow \quad r=r(x,t) 
\ee 
which, introducing the inverse function $x_0(r)$ such that $R(x_0(r))=r$, can also be written as
$x = x_0(r) + 2 \kappa r t$. 
Dividing by $t$ one gets, with $y=x/t$
\be 
r = \frac{y}{2 \kappa} - \frac{1}{2 \kappa t} x_0(r) 
\ee 
This leads to $r_0(y)=\frac{y}{2 \kappa}$ and 
\be 
r_1(y)= - \frac{1}{2 \kappa} x_0\left( \frac{y}{2 \kappa}\right) 
\ee
which shows explicit dependence in the initial condition. Although in
the full problem there is also thermal and
active noise, their role in the bulk of the expanding gas may be subdominant. 
\\

{\it Boundary layer for the expanding phases.} Near the edges of the expanding
gas the above $1/t$ expansion fails and one needs to look for a different scaling
form, as discussed in the text, within a boundary layer of width $\sim \sqrt{t}$.

In the boundary layer we can look for a solution of (\ref{system1_potential_time}-\ref{system2_potential_time}) under the form
\be
r(x,t) = \frac{1}{2} - \frac{1}{\sqrt{t}} \, f(z) \quad , \quad
s(x,t) = \frac{1}{t} \, g(z) \quad , \quad z = \frac{x - (\kappa-a) t}{\sqrt{t}} \;.
\ee
Inserting these forms into \eqref{system2_potential_time} gives at leading order in $\frac{1}{t}$
\be 
g(z) = \frac{v_0}{2\gamma} f'(z) \;.
\ee
Replacing $g(z)$ in \eqref{system1_potential_time}, leads to
\be
\frac{1}{2} f + \frac{z}{2} f' + T_{eff} f'' + 2 \kappa f f' = 0 \quad , \quad T_{eff} = T + \frac{v_0^2}{2\gamma} \;.
\ee
This is exactly the same boundary layer equation as in the passive case with an effective temperature $T_{eff}$. This equation can be made dimensionless
by writing
\be
\tilde f(\tilde z) = \frac{\kappa}{\sqrt{T_{eff}}} f(z) \quad , \quad \tilde z = \frac{z}{\sqrt{T_{eff}}}
\ee
yielding the equation
\be
\frac{1}{2} \tilde f + \frac{\tilde z}{2} \tilde f' + \tilde f'' + 2 \tilde f \tilde f' = 0
\ee
which is solved by
\be
\tilde f (\tilde z) = \frac{e^{-\tilde z^2/4}}{\sqrt{\pi} \, {\rm erfc}(-\tilde z/2)} \;.
\ee
We can thus write the full solution as a function of $\tilde f(\tilde z)$
\be
r(x,t) = \frac{1}{2} - \frac{1}{\kappa} \sqrt{\frac{T_{eff}}{t}} \tilde f(\tilde z) \quad , \quad s(x,t) = \frac{v_0}{2\gamma \kappa t} \tilde f' (\tilde z)  \quad , \quad \tilde z = \frac{x - (\kappa-a) t}{\sqrt{T_{eff} t}}
\ee
and 
\be
\rho_s(x,t) = - \frac{1}{\kappa t} \tilde f'(\tilde z) \quad , \quad \rho_d(x,t) = \frac{v_0}{2\gamma \kappa \sqrt{T_{eff}} \, t^{3/2}} \tilde f'' (\tilde z)
\ee
which is the result that we displayed in the text in the case $a=0$. 


As a final remark let us stress that there is strictly no difference in the above results between the $E_s$ and $E_0$ phases,
which can only be distinguished by looking at spatial scales $O(1)$ to which we now turn.
\\

{\it (ii) Stationary densities for the bound particles.} Contrary to the case $a=0$, we see that a finite fraction $\frac{a}{\kappa}$ of particles remains close to $x=0$. However, because of the rescaling performed above ($ x \sim t$) we do not yet know if these particles form a large cluster at $x=0$ or if they simply have a position which is of order $o(t)$. To answer this question we now focus on particles which have a position $x=O(1)$ and assume that they reach a steady state.
We then need to solve the same system of equations for $r(x)$ and $s(x)$ as in (\ref{system1_lin}-\ref{system2_lin}), but with new boundary conditions $r(\pm \infty)=\frac{a}{2\kappa}$ and $s(\pm \infty)=0$. Integrating \eqref{system1_lin} we obtain for $x > 0$
\be \label{s_function_r_lin3}
s(x) = -\frac{\kappa}{v_0} \big(r(x)^2 - \frac{a^2}{4 \kappa^2} \big) + \frac{a}{v_0} \big(r(x) - \frac{a}{2 \kappa} \big)
= - \frac{\kappa}{v_0} (r - \frac{a}{2 \kappa})^2  \;.
\ee 
Substituting in \eqref{system2_lin} one obtains
\be \label{eqr_potential_confined}
(v_0^2 - (a - 2\kappa r)^2) r' 
= \frac{\gamma \kappa}{2} (\frac{a}{\kappa}- 2 r)^2 \;.
\ee
Let us define $\hat r(x)=\frac{\kappa}{a}r(x)$ and  inject in 
Eq. \eqref{eqr_potential_confined}. One finds that $\hat r(x)$ satisfies the equation for $x>0$
\be \label{eqrhat}
(v_0^2 - a^2(1-2\hat r)^2) \hat r' = \frac{\gamma a}{2}(1 - 2\hat r)^2
\ee 
with $\hat r(\pm \infty)=1/2$. This is exactly the same equation as the one obtained for $r(x)$ in the marginal case, i.e. Eq. \eqref{eqkappa} with $a=\kappa$. This can be understood as follows. Since they are equally split between the two sides, the particles which escape to infinity can be ignored when studying the particles which remain confined in the potential (the force that they exert on a confined particles sums to zero). One then only has to consider $N_{\rm bound}=\frac{a}{\kappa}N$ particles with a repulsive interaction strength $\frac{\kappa}{N}=\frac{a}{N_{\rm bound}}$, i.e. the bound particles behave as if the effective interaction constant was $\kappa_{\rm eff}=a$, corresponding to the critical case described above. 

Thus, using the results above, for $x=O(1)$ and $x>0$, $r(x)$ is given by
\bea  
&& \gamma x = \tilde {\sf f}_a \left(\frac{\kappa}{a} r(x) \right) \quad , \quad a<v_0 \\
&& \gamma x = \tilde {\sf f}_a\left(\frac{\kappa}{a} r(x)\right) - \tilde {\sf f}_a\left(\frac{\kappa}{a} r(0^+)\right) \quad , \quad r(0^+) = \frac{a-v_0}{2\kappa} \quad , \quad a>v_0
\eea  
which correspond respectively to phases $E_s$ and $E_0$,
where $\tilde {\sf f}_a$ is given in \eqref{eqres_potential_akappa}. 
Thus, there is indeed a cluster of particles at $x=0$ only in the case $a>v_0$ (phase $E_0$). 

From this observation one obtains the density $\rho_s(x)$ explicitly in the phases $E_s$ and $E_0$.
It is simply given by the same formulae \eqref{rhosEs} and \eqref{rhosE0} by multiplying
the density by the factor $\frac{a}{\kappa}$. The densities have infinite support and are smooth except at $x=0$. One obtains the 
following behaviors at large $|x|$
\be
\rho_s(x) \simeq \frac{v_0^2}{2 \kappa \gamma x^2} \quad , \quad \rho_d(x) \simeq \frac{v_0^3}{2 \kappa \gamma^2 x^3} 
\ee
so, in that sense, in the phases $E_s$ and $E_0$ the bound part of the gas is always critical (i.e. the decay of the densities are
power laws rather than exponentials as it is in the phases $I_s$ and $I_0$). 

Let us now discuss the behavior of the densities around $x=0$. In the phase $E_s$, applying the results 
obtained above in the phase $I_s$, Eqs. \eqref{r_IsFe_cusp} and \eqref{rhos_IsFe_cusp}, 
to $\hat r(x)$ and performing
the rescaling $r(x)=\frac{a}{\kappa} \hat r(x)$ one finds
\be \label{rhos_Es_cusp}
\rho_s(x) = \frac{a^2}{2 \kappa (v_0^2 - a^2) } \gamma - \frac{a^3 v_0^2}{ 2 \kappa (v_0^2 -a^2)^3 } \gamma^2 |x| 
+ O(x^2) \;.
\ee 
From \eqref{s_function_r_lin3} one finds again that $s'(0^+)= \frac{a}{v_0} r'(0^+) = \frac{a}{v_0} \rho_s(0)$, since
$r(0^+)=0$ in this phase, hence we obtain near $x=0$
\be 
\rho_d(x) \simeq \frac{a^3 \gamma}{2 v_0 \kappa (v_0^2-a^2) } {\rm sgn}(x) \;.
\ee 

In the phase $E_0$ one finds, applying the results obtained above in the phase $I_0$ to $\hat r(x)$ and performing
the rescaling $r(x)=\frac{a}{\kappa} \hat r(x)$ that near $x=0$
\be 
r(x) \simeq \left( r(0^+) + B' \sqrt{|x|} \right)  {\rm sgn}(x)  \quad , \quad \rho_s(x) \simeq \frac{a-v_0}{\kappa}  \delta(x) + \frac{B'}{2 \sqrt{|x|}} \quad {\rm with} \quad B' = \frac{1}{2\kappa} \sqrt{\gamma v_0}  \;, \label{sqrtdiv2} 
\ee 
as well as (using the relation \eqref{s_function_r_lin3}) 
\be
s(x) \simeq -\frac{v_0}{4 \kappa} + B' \sqrt{|x|}  \quad , \quad \rho_d(x) \simeq  \frac{B'}{2 \sqrt{|x|}} \, {\rm sgn}(x) \;.
\ee


\subsection{Passive limit}

In this section we focus on the phase $I_s$ and we consider the passive
limit $v_0 \to +\infty$, $\gamma \to +\infty$ with $T_a=v_0^2/(2 \gamma)$ fixed.
Taking this limit in \eqref{eqres_potential},
one
obtains for $r>0$
\be 
\frac{x}{T_a}=  \frac{1 }{\bar \kappa+a} \log\left( \frac{\bar \kappa + 2 a + 2 \bar \kappa r}{(\bar \kappa + 2 a)(1-2 r) } \right) \;,
\ee 
which leads to, for any $x$ (using that $r(x)$ is odd),
\be  \label{solupassive_r} 
r(x) = \frac{1}{2} \frac{1- e^{- (\bar \kappa + a) |x|/T_a} }{1 + \frac{\bar \kappa}{\bar \kappa + 2 a} e^{- (\bar \kappa + a) |x|/T_a} } {\rm sgn}(x)
\ee 
and thus
\be  \label{solupassive_rho}
\rho(x)= \frac{(\bar \kappa + a)^2}{T (\bar \kappa + 2 a)} \frac{e^{-\frac{(\bar \kappa+a) x}{T}}}{ 
( 1 + \frac{\bar \kappa}{\bar \kappa + 2 a} e^{-\frac{(\bar \kappa+a) x}{T}} )^2} \;.
\ee 
This result can also be obtained by solving directly the Dean-Kawasaki equation for the passive case in the
large $N$ limit, i.e. solving 
\be 
0 = T  r''(x) + ( 2 \bar \kappa r(x)  + a \, {\rm sgn}(x)) r'(x) \;.
\ee 
This solution is valid as long as $\bar \kappa > -a$, i.e. $\kappa<a$.
It is always continuous with infinite support, which is consistent with the phase $I_s$.

The density for the passive system can also be obtained using the Coulomb gas. The $N$ particle equilibrium
Gibbs measure at temperature $T$ is
\be 
\exp( - \frac{\bar \kappa}{N T} \sum_{i<j} |x_i-x_j|- \frac{a}{T} \sum_i |x_i| ) 
\ee  
which is normalizable if $\frac{N-1}{N}  \bar \kappa> -a$.
For large $N$ one can rewrite it in terms of the density, in a Coulomb gas description as
\be 
\exp( - \frac{\bar \kappa N}{2 T} \int dx \rho(x) \rho(x') |x-x'| - \frac{N a}{T}  \int dx \rho(x) |x|  - N \int dx \rho(x) \log \rho(x) ) 
\ee 
with the constraint $\int dx \rho(x)=1$, and we have included the entropy term which, given the way we
scaled the interaction, is of the same order as the other terms.
For $T=O(1)$ and large $N$ the Gibbs measure is concentrated around the 
optimal density, which is obtained as the solution of a self-consistent equation
\be  \label{sc} 
\rho(x) = K e^{- \frac{\bar \kappa}{T} \int dx'  \rho(x') |x-x'| - \frac{a}{T} |x| } \;.
\ee 
One can check that the solution \eqref{solupassive_rho} satisfies this self-consistent equation. Indeed, 
for $x>0$ one has 
\bea  
&& \int dx'  \rho(x') |x-x'| = \int_{-\infty}^0 dx' (r(x')+ \frac{1}{2}) + \int_{0}^x dx' (r(x')+ \frac{1}{2})
- \int_{x}^{+\infty} dx' (r(x')- \frac{1}{2}) \\
&& = - \frac{(\bar \kappa+ 2 a) x}{\bar \kappa} + 2 \frac{T}{\bar \kappa}  \log( \frac{\bar \kappa}{\bar \kappa + 2 a} + e^{\frac{(\bar \kappa+a) x}{T}}) 
\eea   
which shows that \eqref{sc} is indeed obeyed with $K = \frac{(\bar \kappa + a)^2}{T (\bar \kappa + 2 a)} $.
Note that the equivalence of the DK approach and the Coulomb gas in the passive case was discussed
in \cite{PLDRankedDiffusion}.


If $\bar \kappa < -a$, we obtain instead an expanding phase similar to the active case, but where for $x=O(1)$ the density never has a delta peak at $x=0$. Thus in the passive case the phase diagram is much simpler and only exhibits 2 phases, corresponding to $I_s$ and $E_s$ respectively.

\subsection{Moments in the steady states}

For the stationary measures obtained here,here one can always define the reciprocal function $x(r)$ of $r(x)$, 
which has a plateau whenever $r(x)$ has a shock. The moments of $|x|$ for $k \geq 1$ can then be expressed as
\be 
\langle |x|^k \rangle = 2\int_{0}^{+\infty} dx \rho_s(x) x^k =
2 \int_{0^+}^{1/2^-} dr \, x(r)^k \;.
\ee 
In the part of the support where the density is smooth one can use the relation 
$x(r)= \frac{1}{\gamma} {\sf f}_a(r)$ from \eqref{eqres_potential}. Hence in the phases $I_s$ and $I_0$ one has
\be  
\langle |x|^k \rangle = \begin{cases} \frac{2}{\gamma^k} \int_{r(0^+)}^{1/2} dr \, {\sf f}_a(r)^k \quad , \quad \text{in phases} \ I_s \ \text{and} \ I_0
\\
\frac{2}{\gamma^k} \int_{0}^{r(x_e^-)} dr \, {\sf f}_a(r)^k + 2 p_e x_e^k \quad , \quad \text{in phase} \, F_e
\end{cases} 
\ee 
where we recall that $r(x_e^-)=\frac{v_0-a}{\bar \kappa}-\frac{1}{2}$ and $p_e=\frac{\bar \kappa+a-v_0}{\bar \kappa}$ in the phase $F_e$, as well as $r(0^+)=\frac{a-v_0}{2\kappa}$ in $I_0$ and $0$ in $I_s$.

Let us give some explicit formulas in the phase $I_s$. One finds
\be 
\langle |x| \rangle = \frac{\bar \kappa}{2 \gamma} +  \frac{v_0^2 - (\bar \kappa + a)^2}{ \bar \kappa \gamma} \log( \frac{2 (\bar \kappa + a)}{\bar \kappa + 2 a}) \;.
\ee 
Specializing now to the case $a=0$ one obtains
\be 
\langle |x| \rangle = \frac{\bar \kappa}{2\gamma} (1 + (\frac{v_0^2}{\bar \kappa^2} -1) \log 4) \;,
\ee 
as well as
\be \langle x^2 \rangle =  
\frac{\bar \kappa^2}{12 \gamma^2} \left(12   \frac{v_0^2}{\bar \kappa^2}  +\pi ^2
   \left(\frac{v_0^2}{\bar \kappa^2} -1\right)^2-8\right) \;.
\ee 
Clearly the moments of any order are simple polynomials in $v_0$.

\section{More general interaction and self-consistent equation}

In this section we consider more general interactions and derive a self-consistent equation
for the steady state density.

Consider $N$ particles in 1D with positions $x_i$ and total energy 
\be 
E(\vec x) = \frac{1}{N} \sum_{i<j} W(x_i-x_j) + \sum_i V(x_i) 
\ee 
where $V(x)$ is the external potential and where $W(x)$ is the pairwise interaction energy,
the function $W(x)$ being even. 
The equation of motion reads (here we assume no passive noise)
\be 
\frac{dx_i}{dt} = - \frac{1}{N} \sum_{j \neq i} W'(x_i-x_j) - V'(x_i) + v_0 \sigma_i(t) \;.
\ee
Here $W'(x)$ is thus an odd function. Hence either one has $W'(0)=0$ or $W'(x)$ diverges at $x=0$.
In the second case, for divergent repulsive interactions with $W'(0^+)=+\infty$, the particles cannot cross
and we have found 
in \cite{TouzoDBM2023} that the Dean-Kawasaki (DK) method is problematic in the active case.
Hence we will restrict here to the case $W'(0)=0$ where the particles can cross freely.
In that case, the large $N$ limit can be taken more straightforwardly, and 
the method should work, as confirmed by our numerics in the particular case of the rank diffusion, i.e.
$W'(x)= {\rm sgn}(x)$.


Then, the DK method gives, in the limit of large $N$, the evolution equation 
for the densities
\be \label{eqrho1n}
 \partial_t \rho_\sigma(x,t)  =  
\partial_x \left[\rho_\sigma(x,t)  \left( - v_0 \sigma +  V'(x) +
\int dy W'(x-y) (\rho_+(y,t) + \rho_-(y,t) ) \right) \right]  + \gamma \rho_{- \sigma}(x,t) - \gamma \rho_{\sigma}(x,t) \;.
\ee
In terms of the densities $\rho_s$ and $\rho_d$ this leads to 
\bea \label{eqrho2n}
&& \partial_t \rho_s =    \partial_x [ (- v_0 \rho_d - \tilde F(x,t) \rho_s ] \\
&& \partial_t \rho_d =   \partial_x [ (- v_0 \rho_s - \tilde F(x,t) \rho_d ] - 2 \gamma \rho_d\nn
\eea 
where
\be 
\tilde F(x,t) = - V'(x) - \int dy W'(x-y) \rho_s(y,t) \;.
\ee 
Note that the equations \eqref{eqrho2n} are identical to the equations of a single RTP, $N=1$, in a effective time dependent force field
$\tilde F(x,t)$ which depends itself on the time dependent density. One can thus use the known results for that simpler problem,
whenever these are available.

Consider now the stationary measure (which we assume here to exist). The steady state densities must be solution of
\bea \label{eqrho22n}
&& 0 =    \partial_x [ (- v_0 \rho_d - \tilde F(x) \rho_s ] \\
&& 0 =   \partial_x [ (- v_0 \rho_s - \tilde F(x) \rho_d ] - 2 \gamma \rho_d \nn
\eea 
where the effective force field is now static 
\be 
\tilde F(x) = - V'(x) - \int dy  \, W'(x-y) \rho_s(y)
\ee 
and depends on the steady state density. 

These equations can be solved formally since the stationary measure for a single RTP in an arbitrary force field is known. 
Let us recall the main steps. One can integrate the first equation on the real line, leading to
\be 
\rho_d(x) = -\frac{\tilde F(x)}{v_0} \rho_s(x) + C_d  \label{rhodself}
\ee 
where $C_d$ is an integration constant. Assuming now that $\tilde F(x) \rho_s(x)$ as well as $\rho_d(x)$ vanish at infinity
leads to $C_d=0$. Inserting in the first equation we obtain
\be 
2 \gamma \tilde F(x) \rho_s(x)  = \partial_x ( (v_0^2-\tilde F(x)^2) \rho_s(x)) \;.
\ee 
Denoting $\rho_s(x)=f(x)/(v_0^2-\tilde F(x)^2)$, one has $f'(x)=2 \gamma \tilde F(x)/(v_0^2-\tilde F(x)^2) f(x)$.
This equation can be integrated. 
There are then several cases depending on how many roots the equation $\tilde F(x)=v_0^2$. If we assume here
for simplicity the case of a connected support of the density (a single interval which can be infinite)
one obtains
\be 
\rho_s(x) = \frac{K}{v_0^2-\tilde F(x)^2} e^{2 \gamma \int_0^x dz \frac{\tilde F(z)}{v_0^2-\tilde F(z)^2}} \quad , \quad 
\tilde F(z)= F(z) - \int dy \, W'(z-y) \rho_s(y) \label{self}
\ee 
where $K$ is a constant determined by normalization and we have assumed that $x=0$ is within the support of the density.
The equation \eqref{self} must be understood as a self-consistent equation for the steady state density $\rho_s(x)$.
Once $\rho_s(x)$ is known, $\rho_d(x)$ is obtained from \eqref{rhodself}. We defer its general study to future work
and mention here two simple applications. 
\\


{\bf Recovering active rank diffusion}. Let us show how one recovers the solution for
the active rank diffusion problem obtained in our paper. In that case one has $W(x)=\bar \kappa |x|$, 
$W'(x)=\bar \kappa \, {\rm sgn}(x)$, with ${\rm sgn}(0)=0$, and $ \bar \kappa \int dy \, \rho_s(y) {\rm sgn}(x-y)  = 2 \bar \kappa r(x)$. 
Hence the effective force field is $\tilde F(x)=- 2 \bar \kappa r(x)$. 
Considering here for simplicity only the case of the phase $I_s$ where the support of the densities is infinite, 
the self-consistent equation
\eqref{self} then reads
\be \label{sc2} 
r'(x) = \frac{K}{v_0^2-4 \bar \kappa^2 r(x)^2} e^{ -4 \gamma \bar \kappa \int_0^x dz \frac{r(z)}{v_0^2- 4 \bar \kappa^2 r(z)^2}  } \;.
\ee 
It is not so obvious to solve that equation directly. One can guess that there exists a function $g(r)$ such that
\be 
4 \gamma \bar \kappa \int_0^x dz \frac{r(z)}{v_0^2- 4 \bar \kappa^2 r(z)^2} = g(r(x)) \;.
\ee
Note that since $r'(x)$ must vanish at $|x|=+\infty$, the function $g(r)$ must diverge to $+\infty$ at $r=\pm 1/2$.
Taking a derivative w.r.t. $x$ one obtains 
\be 
g'(r(x)) r'(x) = 4 \gamma \bar \kappa  \frac{r(x)}{v_0^2- 4 \bar \kappa^2 r(x)^2} 
\ee 
and using \eqref{sc2} one can close the equation and obtain an equation for $g(r)$
\be 
g'(r) e^{- g(r)} =  \frac{4 \gamma \bar \kappa}{K} r \;.
\ee 
Hence
\be 
e^{-g} =   \frac{2 \gamma \bar \kappa}{K} ( \frac{1}{4}-r^2) 
\ee 
leading, using again  \eqref{sc2} to
\be 
r'(x) = \frac{2 \gamma \bar \kappa}{v_0^2-4 \bar \kappa^2 r(x)^2} ( \frac{1}{4}-r^2)  
\ee 
which is identical to the equation \eqref{rp} of the text. 
\\

{\bf Harmonic interaction}. In the case of an harmonic interaction $W(x)=\frac{\mu}{2} x^2$ ($\mu>0$), we have
\be
\tilde F(z) = - \mu \int dy (z-y)\rho_s(y) \;.
\ee
If we assume that $\rho_s(x)$ is an even function, this becomes
\be
\tilde F(z) = - \mu z \;.
\ee
Therefore the stationary density of particles at large $N$ is exactly the same as for independent particles in a harmonic potential of strength $\mu$ centered at $x=0$ (see e.g. \cite{DKM19})
\be 
\rho_s(x) = \tilde K \left(1-\left(\frac{\mu x}{v_0}\right)^2 \right)^{\frac{\gamma}{\mu}-1} \quad , \quad \tilde K= \frac{2}{4^{\gamma/\mu} B(\gamma/\mu,\gamma/\mu)} \frac{\mu}{v_0} \quad , \quad |x|<v_0/\mu \;.
\ee 
This was to be expected since the sum of the forces of two particles at $\pm x_0$ on a third particle at $x$ is an attractive harmonic force towards $x=0$.
More precisely the equation of motion reads for any finite $N$
\be 
\dot x_i = \frac{\mu}{N} \sum_j (x_j-x_i) + v_0 \sigma_i  = - \mu x_i + \mu \bar x + v_0 \sigma_i \quad , \quad \bar x = \frac{1}{N} \sum_j x_j
\ee 
and in the large $N$ limit the position of the center of mass $\bar x$ tends to zero, leading to an effective decoupling.

For finite $N$, it is more convenient to go to the reference frame of the center of mass
(which, as in the rank interaction case follows a simple diffusion with $D_N=\frac{1}{N}(T + \frac{v_0^2}{2 \gamma})$
and define $\tilde x_i=x_i-\bar x$. Using that $\frac{d}{dt} \bar x= \frac{v_0}{N} \sum_i \sigma_i(t)$ 
it satisfies 
\be \label{eq_harmonic_finiteN}
\frac{d\tilde x_i}{dt} = \frac{d x_i}{dt} - \frac{d\bar x}{dt} = - \mu \tilde x_i + \big( 1-\frac{1}{N} \big) v_0 \sigma_i(t) - \frac{v_0}{N} \sum_{j(\neq i)} \sigma_j(t) \;.
\ee
For any value of $N$, the equation \eqref{eq_harmonic_finiteN} can be solved explicitly, 
and from it one can compute the first and second moments. One finds
$\langle \tilde x_i(t) \rangle = \tilde x_i(0) e^{-\mu t}$ and
\bea 
\langle \tilde x_i(t)^2 \rangle - \langle \tilde x_i(t) \rangle^2 &=& \big( 1-\frac{1}{N} \big) v_0^2 \left( \frac{1}{2\gamma \mu + \mu^2} + \frac{2 e^{-(\mu+2\gamma)t}}{4\gamma^2-\mu^2} + \frac{e^{-2\mu t}}{\mu(\mu-2\gamma)} \right) \quad \xrightarrow[t \to +\infty]{}  ( 1-\frac{1}{N})  \frac{v_0^2}{\mu(2 \gamma + \mu)} \quad 
{\rm for} \, \mu \neq 2 \gamma \nonumber \\
&=& \big( 1-\frac{1}{N} \big) \frac{v_0^2}{8\gamma^2} \left( 1 - e^{-4\gamma t} - 4 \gamma t e^{-4\gamma t} \right) 
\xrightarrow[t \to +\infty]{} ( 1-\frac{1}{N}) \frac{v_0^2}{8\gamma^2} \quad {\rm for} \, \mu = 2 \gamma
\eea
where the average includes an average over the initial $\sigma_i(0)=\pm 1$, and we have used that
$\langle \sigma_i(t) \sigma_j(t') \rangle=e^{-2 \gamma|t-t'|} \delta_{ij}$. 
Hence, apart from the factor $1-\frac{1}{N}$, it is identical to the moment for 
a single RTP in a quadratic well \cite{DKM19}.

\end{widetext}


\begin{thebibliography}{}
\bibitem{footnoteparity}
Since f(r) is odd, this is the same as imposing $r(0)=0$.
\bibitem{footnoter1} Note however that the higher order terms, e.g. $r_1(y)$,
depend on the initial condition \cite{SM}.
\bibitem{footnoteC}
Solutions with $0< C < 1$, with a shock at $x=0$ are in principle possible but 
are excluded by our numerical results.
\end{thebibliography}
\end{document}